\documentclass[twocolumn,english,aps,physrev,floatfix,amssymb,superscriptaddress,a4paper]{revtex4-1}
\usepackage{lipsum}
\usepackage{amsmath}
\usepackage{amssymb}
\usepackage{graphicx}
\usepackage{physics}
\usepackage{hyperref}
\usepackage[dvipsnames]{xcolor}
\usepackage{mathtools}
\usepackage{soul}

\usepackage[caption=false]{subfig}
\hypersetup{
	colorlinks=true,
	hypertexnames=true
}

\newcommand{\Hmax}{H_{\mathrm{max}}}
\newcommand{\HsL}{H_{s,{\scriptscriptstyle L}}}
\newcommand{\HsR}{H_{s,{\scriptscriptstyle R}}}
\newcommand{\HpL}{H_{p,{\scriptscriptstyle L}}}
\newcommand{\HpR}{H_{p,{\scriptscriptstyle R}}}
\newcommand{\isL}{i_{s,{\scriptscriptstyle L}}}
\newcommand{\isR}{i_{s,{\scriptscriptstyle R}}}
\newcommand{\ipL}{i_{p,{\scriptscriptstyle L}}}
\newcommand{\ipR}{i_{p,{\scriptscriptstyle R}}}

\newcommand{\Hast}{H^\ast}
\newcommand{\HastL}{H^\ast_{\scriptscriptstyle L}}
\newcommand{\HastR}{H^\ast_{\scriptscriptstyle R}}

\begin{document}

\title{Superconductivity in atomically thin films: 2D critical state model}

\author{Filippo Gaggioli}
\affiliation
{Institut f\"ur Theoretische Physik, ETH Z\"urich,
CH-8093 Z\"urich, Switzerland}

\author{Gianni Blatter}
\affiliation
{Institut f\"ur Theoretische Physik, ETH Z\"urich,
CH-8093 Z\"urich, Switzerland}

\author{Kostya S. Novoselov}
\affiliation
{Institute for Functional Intelligent Materials, 
National University of Singapore, 117544, Singapore}

\author{Vadim B. Geshkenbein}
\affiliation
{Institut f\"ur Theoretische Physik, ETH Z\"urich,
CH-8093 Z\"urich, Switzerland}

\date{\today}

\begin{abstract}
The comprehensive understanding of superconductivity is a multi-scale task
that involves several levels, starting from the electronic scale determining
the microscopic mechanism, going to the phenomenological scale describing
vortices and the continuum-elastic scale describing vortex matter, to the
macroscopic scale relevant in technological applications. The prime example
for such a macro-phenomenological description is the Bean model that is hugely
successful in describing the magnetic and transport properties of bulk
superconducting devices. Motivated by the development of novel devices based
on superconductivity in atomically thin films, such as twisted-layer graphene,
here, we present a simple macro-phenomenological description of the critical
state in such two-dimensional (2D) thin films. While transverse screening and
demagnetization can be neglected in these systems, thereby simplifying the
task in comparison with usual film- and platelet shaped samples, surface- and
bulk pinning are important elements to be included. We use our 2D critical
state model to describe the transport and magnetic properties of 2D thin-film
devices, including the phenomenon of non-reciprocal transport in devices with
asymmetric boundaries and the superconducting diode effect.
\end{abstract}

\maketitle


\section{Introduction}

Technologically relevant superconductors are usually type II materials, with
vortices determining their phenomenological properties \cite{Abrikosov_1957}.
In technical applications, the material is often described through its
macro-phenomenological properties where vortices appear as an effective medium
rather than individual objects.  The first such description was the Bean model
\cite{Bean_1962} that describes magnetic- and transport properties of bulk
material in terms of the critical current density $j_p(B)$ due to bulk pinning
\cite{LarkinOvch_1979}, the latter taking the role of a constitutive material
relation (here, $B$ denotes the magnetic induction that is related to the
vortex density $n = 1/a_0^2 = B/\Phi_0$, $\Phi_0 = hc/2e$ is the
superconducting flux quantum, and $a_0$ the vortex separation). A second type
of macro-phenomenological description applies to flat samples, e.g., 
films and platelet-shaped samples as they appeared with the then-novel cuprate
superconductors in the late eighties and nineties.  Besides bulk pinning
\cite{LarkinOvch_1979}, surface pinning \cite{Bean_1964,Clem_1974,
Benkraouda_1998,Blatter_2008}, geometric barriers \cite{Zeldov_1994_geom,
Benkraouda_1996,Brandt_1999, Blatter_2008}, and demagnetization phenomena
\cite{Osborn_1945,Brandt_1993, Zeldov_1994,Blatter_2008} become relevant.
Today, much interest is focused on atomically thin van der Waals materials
\cite{Geim_2013,Novoselov_2016} with superconducting properties
\cite{Saito_2016}, as found in transition metal dichalcogenides \cite{Lu_2015,
Xi_2016} and twisted layer graphene \cite{Cao_2018_MATBG, Yankowitz_2019}.
While significant efforts have focused on explaining the microscopic aspects
of superconductivity in such atomically thin materials \cite{Saito_2016,
Qiu_2021, Das_2023, Balents_2020}, less emphasis has been devoted to studying
the phenomenology of these systems. In the present paper, we introduce a third
kind of macro-phenomenological model that describes the two-dimensional (2D)
critical state of atomically thin superconductors and use it to determine the
relevant transport- and magnetic properties of these materials.

Atomically thin superconductors can easily be tuned, a feature owed to their
nanometric thickness.  E.g., the surface barrier can be manipulated not only
via the geometric design of the edges \cite{Clem_2011}, but also via a finite
gate voltage \cite{Rocci_2021}.  In a recent work, the interplay between
vortex pinning, the moir\'e pattern in $\text{NbSe}_2$-layers, and the effects
of a perpendicular electric field has been studied \cite{Gaggioli_2023_novo},
demonstrating the expanded toolkit for engineering superconductivity in this
type of materials.  This high degree of tunability makes 2D superconductors an
ideal platform for realizing interesting and novel devices, see for example
the numerous reports on the so-called superconducting diode effect
\cite{Vodolazov_2005,Diez-Merida_2021,Lyu_2021,Pal_2021, Baumgartner_2022,
Bauriedl_2022, Golod_2022,Fominov_2022,Lin_2022,Suri_2022,Gutfreund_2023,
Margineda_2023, Sundaresh_2023, Hou_2023,Paolucci_2023,Nakamura_2023,
Nadeem_2023, Gaggioli_2023_novo}.  One important goal of this work then is to
provide a simple macro-phenomenological model capable of describing the
properties of these 2D materials and their physical performance in technical
applications.

\begin{figure}[!ht]
	\centering
        \includegraphics[width = 1.\columnwidth]{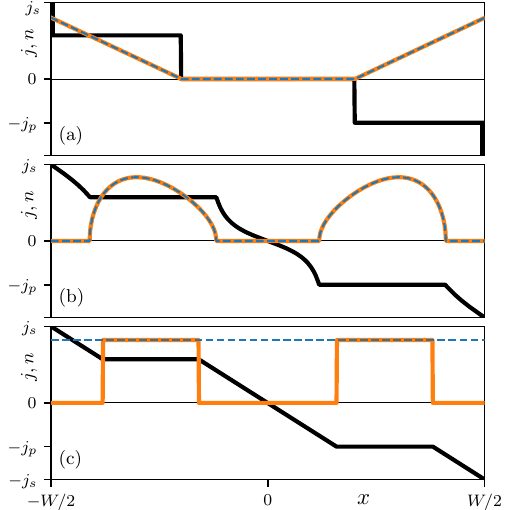}
	\caption[Comparison of critical states] {Illustration of the critical
	state (partial penetration) for three types of sample geometries, bulk
	(a), flat samples (b), and 2D thin films (c), in a (perpendicular)
	magnetic field $\mathbf{H} \parallel \mathbf{z}$.  Current- and vortex
	densities $j(x)$ and $n(x)$ are shown as black and orange lines; the
	corresponding field distribution $B(x)$ is shown as a blue dashed
	line.  (a) In the Bean model \cite{Bean_1962}, the current density is
	constant and equal to $\pm j_p$ close to the edges where vortices are
	trapped, while $n(x)$ decreases linearly with slope $\propto \mp j_p$
	inside the sample; below full penetration, both densities vanish in
	the centre.  Surface (i.e., Meissner) currents flow on a scale $\sim
	a_0^2/\xi > a_0$ at the boundaries \cite{Clem_1974,Blatter_2008}. (b)
	In a flat sample, e.g., a film- or platelet shaped superconductor,
	demagnetization effects are strong and Meissner currents flow along
	all of the sample boundaries \cite{Brandt_1993,Zeldov_1994}; upon
	overcoming the surface- and geometrical barriers at the sample
	boundaries, vortices are driven to the inside where they assume a pair
	of dome-shaped density profiles \cite{Zeldov_1994_geom} that is
	determined (in position and shape) by the pinning current density
	$j_p$.  (c) In 2D thin-film superconductors, screening can be
	neglected, the magnetic field penetrates homogeneously, with a
	constant $B \approx H$ everywhere.  The current distribution follows
	from the Maxwell-London equation \eqref{eq:maxwell_london}; it
	displays a linear slope $\propto -H$ and assumes large values up to
	the surface critical current density $j_s$.  Vortices enter the film
	when the current density $j$ at the edges reaches the surface critical
	value $\pm j_s$.  Two symmetric plateaus with $j=\pm j_p$ mark the
	regions where vortices with density $n(x) = B/\Phi_0$ have penetrated
	and become pinned.  At partial penetration, the center of the sample
	is vortex-free with the current density again changing linearly with
	slope $\propto -H$. While in (a), the bulk critical current density
	$j_p$ is the dominating parameter, it is the surface critical current
	density $j_s$ that dominates the critical state in (c).
}
    \label{fig:crit_state}
\end{figure}

The Bean model has been hugely successfull in explaining the magnetic- and
transport properties of the mixed state in bulk material.  It reduces the
complex pinning behavior of the vortex matter to a simple constitutive
material relation $j_p(B)$. In its simplest (original) variant, $j_p$ is
assumed to be independent on the $B$-field (i.e., the vortex density $n$; we
assume $\mathbf{B} \parallel \mathbf{z}$) and solving Maxwell-Amp\`ere's
equation $\nabla \wedge \mathbf{B} = (4\pi/c) \,\mathbf{j}_p$ provides the
magnetic response of a bulk sample as shown in Fig.\ \ref{fig:crit_state}(a),
i.e., the current- and vortex densities $j(x)$ and $n(x) = B(x)/\Phi_0$ across
the superconductor (we assume a finite-width sample along $x$ and an infinite
extension along $y$).  The simplest critical state is characterized by a
constant current density $j = j_p$ in the penetrated part of the sample that
is associated with a linear decrease in (pinned) vortex density $n(x)$ away
from the sample surface, while both current- and vortex densities vanish
inside.  Increasing the external field $H$, the vortices penetrate to the
middle of the sample (state of full penetration) and the screening saturates
with the magnetization $M = -j_p W/4c$.  Accounting for surface pinning, the
vortex entry into the sample is delayed \cite{Clem_1974} and additional layers
of Meissner currents flow at the edges, as indicated in Fig.\
\ref{fig:crit_state}(a).

With the emergence of high-temperature superconductors, numerous studies
have focused on the phenomenological properties arising from their
platelet/film geometry with typical film thickness $d$ larger than the London
penetration depth $\lambda$. We separate such `flat samples' with $d \gg
\lambda$ from the `two-dimensional (2D) thin films' with $d \ll \lambda$, the
geometry we will focus on in the present paper, see below.  Placing a flat
sample ($d \gg \lambda$) in a perpendicular magnetic field, i.e., with the
shortest dimension $d$ parallel to the external field $\mathbf{H} \parallel
\mathbf{z}$, demagnetization effects are strong and the field lines bend
around the sample, resulting in Meissner screening currents flowing along all
sample boundaries including the top and bottom surfaces perpendicular to the
field \cite{Brandt_1993,Zeldov_1994}. Furthermore, surface- and geometric edge
effects \cite{Clem_1974,Zeldov_1994_geom,Benkraouda_1996,
Benkraouda_1998,Blatter_2008} determine the field where vortices penetrate
into the sample.  In figure \ref{fig:crit_state}(b), we illustrate the
resulting critical state in flat samples involving Meissner- and
vortex-related current densities: since Meissner currents flow on all sample
boundaries including the top and bottom surfaces \cite{Brandt_1993,
Zeldov_1994, Zeldov_1994_geom}, vortices are pushed into the sample as soon as
the surface/geometrical barrier is overcome; they form two isolated vortex
domes where currents $j = \pm j_p$ flow due to bulk pinning.  Meissner
currents outside these domes seemingly diverge at the film boundaries (they
are cut off at distance $d$ from the edge) and connect the
vortex-pinning-induced current plateaus across the film center. The magnetic
field in $z$ direction (evaluated in the film middle at $z=0$) vanishes
everywhere due to Meissner screening currents, except for the inside of vortex
domes and the thin $\lambda$ layer at the edges (note the difference between
the domes in Fig.\ \ref{fig:crit_state}(b) and the corresponding results in
Refs.\ \onlinecite{Brandt_1993,Zeldov_1994} which is due to the missing of the
geometrical barrier in this early work on flat samples: the large Meissner
currents at the edges drive vortices, that have overcome the geometrical
barrier, to the inside of the film).  Hence, a rich critical-state
phenomenology (involving conformal mappings and theory of analytic functions
in the underlying mathematics \cite{Swan_1968,Huebener_1972,
Zeldov_1994_geom,Blatter_2008}) characterizes the analysis of flat
superconductors that goes beyond the simple description of bulk pinning within
the Bean model.

The critical state in 2D thin films with $d \ll \lambda$ is characterized by
the weakness of superconducting screening: as the film thickness $d$ drops
below the bulk penetration depth $\lambda$, the effective screening length is
given by the Pearl length $\lambda_\perp = 2 \lambda^2/d$ \cite{Pearl_1964}.
For thicknesses $d$ in the nanometric range, typical of atomically thin
devices, the effective penetration depth $\lambda_\perp$ is much larger than
the coherence length $\xi$, bringing these materials under the type II
paradigm. Even more, $\lambda_\perp$ easily exceeds the film width $W$,
$\lambda_\perp \gg  W$, and the magnetic field remains unscreened, with $B
\approx H$ throughout the film width. This then defines the parametric
regime that we focus on in the present paper: 2D thin films with thickness $d
\ll \lambda$ and a finite width $W$ such that $\lambda_\perp \gg W$. This results
in the absence of screening and demagnetization effects and allows us to
formulate a simple theory for the critical state of such 2D thin-film material
that departs significantly from its counterpart in bulk and flat samples, see
Fig.\ \ref{fig:crit_state}(c).

While demagnetization and screening can be ignored in 2D thin films with $d
\ll \lambda$, both surface- and bulk pinning effects play important roles and
we will briefly discuss both, one after the other. The vanishing of the
current density $j_\perp$ at the film boundary generates a {\it surface
barrier} for vortex entry (conveniently described by a fictitious image vortex
of opposite circulation) that allows for a surface current density $j_s$ of
order of the depairing current density $j_0\equiv c \Phi_0/
12\sqrt{3}\pi^2\lambda^2\xi$ to flow at a distance $\lambda_\perp$ from the
boundaries, i.e., everywhere inside the sample in a typical situation where
$\lambda_\perp \gg W$.  Combining Maxwell-Amp\`ere's and London's equation
$\mathbf{j} = -(c/2\pi\lambda_\perp)\, \mathbf{A}$ (where $\mathbf{A}$ is the
vector potential associated with the induction $\mathbf{B}$) then tells us
that the current changes linearly $\propto -H$ inside the film as long as $H$
remains small, $H < H_s \propto j_s$. Pushing the external field $H$ beyond
the surface penetration field $H_s$ drives the current density $j$ at the
boundaries beyond $j_s$ and vortices enter the film. These are driven towards
the film center where they accumulate in the shape of a rectangular box, a
'plateau' replacing the `dome' in flat samples \cite{Zeldov_1994_geom} of
width $d \gg \lambda$. As the vortex box expands with further increasing
magnetic field $H$, the surface current is squeezed to a narrow layer that
reaches the width $\xi$ at the scale of the upper critical field $H_{c2}$.

In the presence of {\it `bulk' pinning} with a finite critical current density
$j_p >0$, vortices are driven towards the thin-film center as long as the
local current density $j$ stays larger then $j_p$. Hence, a symmetric pair of
vortex boxes appear away from the boundaries within the 2D thin film as
illustrated in Fig.\ \ref{fig:crit_state}(c); they expand with increasing
drive $H$ and reach the film center (to scale $a_0$) at the full-penetration
field $\sim (j_p/j_0)^2 H_{c2}$.  As a result, the critical state in 2D
thin films differs considerably from those in bulk and flat samples.
Particularly noteworthy is the exchange of constant and linear dependencies in
the field- and current traces when going between the bulk and the 2D critical
states, see Figs.\ \ref{fig:crit_state}(a) and (c), while the comparison
between the critical states of flat samples and 2D thin films, see Figs.\
\ref{fig:crit_state}(b) and (c), is of a more quantitative nature. Note that a
similar absence of screening characterizes the situation of a thin film with
$d < \lambda$ in the parallel-field geometry, see Refs.\
\onlinecite{Shmidt_1974, Gurevich_1994}.

\begin{figure}
        \centering
        \includegraphics[width = 1.\columnwidth]{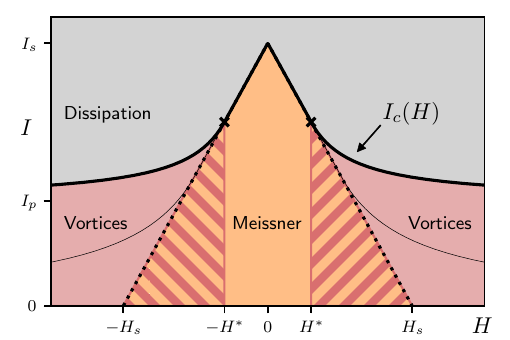}
	\caption[Hysteretic phase diagram]{Critical current $I_c(H)$ (black
	solid line) and hysteretic diagram (colors) for a 2D thin-film
	superconductor in a perpendicular magnetic field $H$.  The critical
	current $I_c(H)$ provides the maximal dissipation-free current
	supported by the superconducting film.  At small fields $|H|\leq
	\Hast$, the critical current is determined by the surface pinning
	inhibiting vortex entry at the sample boundary, with $I_c(|H|\leq
	\Hast)$ decreasing linearly, see \eqref{eq:I_c_meissner_state}. For
	fields beyond $\Hast$, the critical current $I_c(|H| > \Hast)$ is
	limited by vortex motion and decays $\propto 1/|H|$ before saturating
	at the bulk pinning current $I_p$, see \eqref{eq:I_c_mixed_state}.
	Upon increasing the current from zero at fixed $H$, vortices enter the
	film for currents $I > I_s(H) = I_s(1-|H|/H_s)$ (black dotted line,
	see \eqref{eq:I_s_vs_H}) that delimits the Meissner state (orange) on
	increasing currents.  Decreasing the current below $I_c(H)$ at fixed
	$|H| > \Hast$, the film enters the mixed state (red and red stripes).
	In the absence of bulk pinning, $\Hast = H_s/2$ and $I_c(|H| > \Hast)
	= I_s H_s/4 |H|$ (thin solid line).}
    \label{fig:phase_diagram}
\end{figure}

Extending the above critical-state analysis in 2D thin films to include
transport currents $I$, we can calculate the critical current $I_c(H)$ of
devices in the presence of magnetic fields $H$ as shown in Fig.\
\ref{fig:phase_diagram}. We find that the large currents associated with the
surface barrier dominate the current flow and determine the linear
field-dependence \cite{Maksimova_1998,Plourde_2001,Vodolazov_2005} of the
critical current $I_c(H) - I_c(0) \propto |H|$ at small magnetic fields
$|H|\leq H^\ast$ before vortex entry ($H^\ast = H_s/2$ in the absence of bulk
pinning), see the solid black line in Fig.\ \ref{fig:phase_diagram}.  At
larger fields, vortices occupy a large part of the sample and $I_c(H)$
decreases nonlinearly to a saturation value determined by the bulk critical
current density $j_p$.  The figure also informs about the different phases
that appear in 2D thin films at given values of field $H$ and
current $I$, dissipative (grey), Meissner (orange), and mixed (red); the
latter two, Meissner- versus mixed states depend on the preparation of the
state, here shown is the result of increasing (decreasing) the current $I$
from zero (from $I_c(H)$) at fixed $H$.  A further extension of the model to
account for unequal surfaces in asymmetric devices leads us to non-reciprocal
transport that is the basis for the superconducting diode effect.

In this paper, we build on previous studies \cite{Maksimova_1998,Plourde_2001,
Vodolazov_2005} to provide an exhaustive description of the 2D critical state
model of 2D thin-film superconductors with thickness $d \ll \lambda$.  In
Sec.\ \ref{sec:maxwell_london}, we introduce the 2D critical state model for
such 2D thin films based on the Maxwell-London equation that governs the
current distribution across such films.  We work in the limit $W\ll
\lambda_\perp$ typical of atomically-thin strips where we can neglect effects
of screening, while fully accounting for both surface- and bulk vortex
pinning.  We make use of this 2D critical state model to calculate vortex- as
well as current-density profiles, first in a finite field without transport,
see Sec.\ \ref{sec:no-transport}, that leads us to Fig.\
\ref{fig:crit_state}(c).  In Sec.\ \ref{sec:critical_current}, we determine
the field $H$ dependence of the critical current $I_c(H)$, see Fig.\
\ref{fig:phase_diagram}, and then find the critical state for a transport
current $0 < I < I_c(H)$ in between.  In Sec.\ \ref{sec:asymmetric_films}, we
determine the non-reciprocal transport properties of asymmetric devices and
discuss the superconducting diode effect.  We find (in Sec.\
\ref{sec:asymm_Ic}) that the critical current $I_c(H)$ reaches its maximum at
a finite peak field $\Hmax$. The different critical currents for positive and
negative currents then give rise to the superconducting diode effect.
Including bulk pinning, we find an upper bound on the diode efficiency
$\eta(H)$ of the 2D thin film even in strongly asymmetric devices, see Sec.\
\ref{sec:diode_efficiency}. In Sec.\ \ref{sec:sc_diode_exp}, we compare our
critical state predictions with recent experiments.  While a great deal of
transport measurements have been reported on atomically thin superconductors,
not much attention has been paid to the magnetic properties of these
materials.  In the final part of this paper, see Sec.\ \ref{sec:magnetic}, we
use our 2D critical state model to determine the hysteretic moment $m(H)$ that
exhibits numerous features generated by the interplay of surface and bulk
pinning. Besides strong experimental signatures, such as kinks, we find a rich
variety of vortex configurations such as vortex--anti-vortex coexistence
fronts inside the sample. Section \ref{sec:film_conclusions} summarizes our
results and provides some conclusions.

\section{2D critical state model}\label{sec:maxwell_london}

We consider a thin ($d \ll \lambda$) superconducting film (in the $x-y$ plane)
in a perpendicular magnetic field $\mathbf{H} = H \hat{\mathbf{z}}$ and
introduce a new critical state model to determine the relevant transport and
magnetic responses.  Inside the superconductor, the current density
distribution $\mathbf{j}$ is determined self-consistently by the interplay of
the external field $\mathbf{H}$, the distribution of superconducting vortices
and the self-field $\mathbf{B}_\mathrm{self}$ induced by the current density
$\mathbf{j}$, as given by the Biot-Savart integral. Our 2D thin film has a
finite width $-W/2\leq x\leq W/2$ along $x$ and extends to infinity along $y$.
For a film of thickness $d$ smaller than the London length $\lambda$, the
current density $\mathbf{j}(\mathbf{R})$ is homogeneous in the field direction
$z$ and we define the sheet current density $\mathbf{i}(\mathbf{R}) \equiv
\mathbf{j}(\mathbf{R}) d$ that depends on the position $\mathbf{R}$ within the
plane of the film.  The sheet current density $\mathbf{i}(\mathbf{R})$ is
related to the electromagnetic vector potential $\mathbf{A}(\mathbf{r})$ and
the condensate phase $\theta(\mathbf{R})$ through the London equation
\begin{equation}\label{eq:london_original}
   \mathbf{i}(\mathbf{R})=-\frac{cd}{4\pi\lambda^2}\left[\mathbf{A}(\mathbf{R},z=0)
   - \frac{\Phi_0}{2\pi}\mathbf{\nabla} \theta(\mathbf{R})\right].
\end{equation}
We take the curl of Eq.\ \eqref{eq:london_original} and rewrite the total
field $\mathbf{B} = \nabla\times \mathbf{A}$ as the sum of the external- and
Biot-Savart fields $\mathbf{H}$ and $\mathbf{B}_\mathrm{self}(\mathbf{r})$;
furthermore, we replace the singular gauge part $\nabla \times
[\nabla \theta(\mathbf{R})] = 2 \pi \hat{\mathbf{z}} \sum_n \delta
(\mathbf{R}-\mathbf{R}_n)$ by the vortex density $\hat{\mathbf{z}}
n(\mathbf{R})$. Accounting for the film geometry and boundary conditions
(infinite extension and current along $y$), the $z$-component of the curl
provides us with the Maxwell-London equation \cite{Larkin_1972} for the sheet
current density $i(x) \equiv i_y(x)$
\begin{equation}\label{eq:maxwell_london_full}
   \frac{\mathrm{d}\,i}{\mathrm{d}\,x}
   = - \frac{cd}{4\pi\lambda^2}\left[H - n(x)\Phi_0\ + B_\mathrm{self}(x)\right],
\end{equation}
with the $z$-component of the self-field
\begin{equation}\label{eq:self_field}
   B_\mathrm{self}(x) = \frac{2}{c}\int_{-W/2}^{W/2}
   \frac{i(x')\,\mathrm{d}x'}{x'-x}.
\end{equation}
Equation \eqref{eq:maxwell_london_full} defines an integro-differential
equation for $i(x)$ with a characteristic length scale $\lambda_\perp =
2\lambda^2/d$ that follows from the comparison of the differential $\partial_x
i$ with the self-field term $\sim - (d/2\pi \lambda^2) i(x)
\ln[(W/2-x)/(W/2+x)]$ on the right hand side.  Estimating the self-field
\eqref{eq:self_field} from the current in the absence of screening, see Eqs.\
\eqref{eq:maxwell_london} and \eqref{eq:london} below, we find that
$B_\mathrm{self} \sim - (W/\lambda_\perp) H$.  For 2D superconducting strips
with $W \ll \lambda_\perp$, the self-field is negligible and the total
magnetic field inside and outside the material has the same value $B \approx
H$.  The equation for the current distribution inside such 2D superconducting
thin films therefore simplifies to
\begin{equation}\label{eq:maxwell_london}
   \frac{\mathrm{d}\,i}{\mathrm{d}\,x}\approx -\frac{cd}{4\pi\lambda^2}
   \left[H - n(x)\Phi_0 \right],
\end{equation}
a simple first-order differential equation with a source term that depends on
the vortex density $n(x)$.

In the following, we will integrate Eq.\ \eqref{eq:maxwell_london} to find the
current-density profile at given field $H$ and fixed bias current $I=\int\!
i(x)\,dx$ as summarized in Figs.\ \ref{fig:current_density}(b)--(e) and
\ref{fig:current_density_rev}(b).  In doing so, we will determine the critical
current $I_c(H)$ as shown in Fig.\ \ref{fig:phase_diagram} and characterize
the (hysteretic) critical state of thin superconductors with $d \ll \lambda$.

\subsection{Without bias current $I = 0$}\label{sec:no-transport}

To start with, we consider Eq.\ \eqref{eq:maxwell_london} in the absence of
external bias currents, $I=0$.  In the Meissner state, i.e., when $n(x)=0$,
the current-density profile is linear,
\begin{equation}\label{eq:london}
   i(x) = -\frac{c\,d H}{4\pi\lambda^2}x,
\end{equation}
and takes opposite values on the two sides of the 2D film.

Upon increasing the magnetic field $H$, so-called Pearl vortices start
penetrating the superconductor; the latter are disk-like objects with
supercurrents circulating on the scale $\lambda_\perp$ that produce a magnetic
field of monopolar shape in the upper/lower half-plane, in contrast to the
usual flux-tube associated with the Abrikosov vortex in a bulk sample.  The
nucleation of vortices from the edges of the superconducting film is hampered
by the presence of a steep surface barrier. In the ideal scenario, the latter
can be overcome by a transport- or field-induced current density $j$ when this
approaches the depairing current density $j_0$ at the boundary.  Indeed, in
the narrow 2D film ($W \ll \lambda_\perp$) the self-field can be neglected and
the current distribution (at zero field) is homogeneous and equivalent to that
in a 1D wire \cite{Tinkham_2004}, which results in the criterion $j_s = j_0$
for the critical surface current density $j_s$.  In real materials, the
critical current density $j_s$ where vortices start penetrating inside the
superconductor is typically smaller than the depairing value \cite{Xu_2010}
due to edge roughness, imperfections, or voltage gating \cite{Rocci_2021}.

In the absence of an external bias current $I = 0$, vortices penetrate inside
the sample when the magnetic field reaches the surface penetration value
$H_s$, i.e., when the magnitude of the current density $i(x)$ at the edges
$x=\pm W/2$ reaches the critical value $i_s\equiv j_s d$. Using Eq.\
\eqref{eq:london}, the condition $i_s = [(cd\,H_s)/(4\pi\lambda^2)](W/2)$
provides us with the surface penetration field
\begin{equation}\label{eq:H_s}
   H_s\equiv\frac{8\pi}{c\,d}\frac{\lambda^2}{W}\,i_s
\end{equation}
where the superconductor enters the mixed state for $I=0$.  Using the above
criterion $j_s = j_0$ for the surface current density, the vortex penetration
field \eqref{eq:H_s} reads
\begin{equation}\label{eq:penetration_field}
   H_s = \frac{2\Phi_0}{3\sqrt{3}\pi\xi W} 
   \sim \frac{\Phi_0}{\xi W} \sim \frac{\lambda}{W}H_c,
\end{equation}
where the last estimate relates $H_s$ to the thermodynamic critical field $H_c
\equiv \Phi_0/2\sqrt{2}\pi\lambda\xi$ that sets the magnetic scale for the
penetration of vortices across the surface of a bulk sample in a parallel
field \cite{Bean_1964} (in Ref.\ \onlinecite{Maksimova_1998}, another
criterion was used, that resulted in a smaller penetration field $H_s =
\Phi_0/2\pi\xi W$, see also Refs.\ \onlinecite{Shmidt_1974, Gurevich_1994}
where the same criterion gives $H_s = \Phi_0/2\pi \xi d$ for a film in
parallel field).  Indeed, vortices penetrate inside the thin film when
currents of order $j_0$ flow across the full width $W$; this has to be
contrasted with the case of a bulk sample where large currents of order $j_0$
are limited to the London length $\lambda$, what explains the factor
$\lambda/W$ in Eq.\ \eqref{eq:penetration_field}.

In the absence of bulk pinning, the vortices that penetrate inside the 2D film
accumulate in a finite region around the center of the superconductor where
$i(x) = 0$.  In this region, the derivative $\mathrm{d}i/\mathrm{d}x$ in Eq.\
\eqref{eq:maxwell_london} vanishes, implying that the density is constant and
given by $n = H/\Phi_0$; this `vortex box' replaces the `vortex dome' in the
flat sample.  The solution of \eqref{eq:maxwell_london} then consists of
piecewise linear parts $i(x) = \mp [i_s - (cd\,H/4\pi\lambda^2)(W/2 \mp x)]$
near the vortex-free right and left edges, respectively, and a vanishing
current $i(x) = 0$ in the vortex box; the two solutions merge at the box edges
$\pm x_v(H)$ that derive from the condition $i(\pm x_v) = 0$ imposed on the
linear part of the solution,
\begin{equation}\label{eq:x_0_symm}
   x_v(|H| > H_s) = \frac{W}{2}\left(1 - \frac{H_s}{|H|}\right).
\end{equation}
The resulting current-density profile is shown as a red line in the inset
of Fig.\ \ref{fig:current_density}(e) further below.

At finite bulk pinning $j_p>0$ and large enough fields $|H| > H_s$, vortices
from both sides are driven towards the center of the film as long as the local
current density $i(x)$ stays larger than $i_p$.  Hence, two symmetric vortex
plateaus $|x|\in\left[x_i(H),x_o(H)\right]$ appear away from the boundaries,
see Fig.\ \ref{fig:crit_state}(c) in the introduction.  The current density is
constant and equal to $i_p$ inside these regions and thus the slope $di/dx$ in
\eqref{eq:maxwell_london} vanishes, again producing the vortex density
$n(x)=H/\Phi_0$.  Matching the constant current $i=i_p$ within the plateaus to
the (properly shifted) linear current densities \eqref{eq:london} in the
vortex-free regions, we find the expressions for the inner and outer
boundaries $x_i(H)$ and $x_o(H)$, 	
\begin{align}\label{eq:x_m_x_M}
   \left(x_i,x_o\right)(|H|> H_s)= \frac{W}{2}
   \left(\frac{H_p}{|H|},\,1-\frac{H_s - H_p}{|H|}\right),
\end{align}
with
\begin{equation}\label{eq:H_p} 
   H_p \equiv\frac{8\pi}{c\,d}\frac{\lambda^2}{W}\,i_p
\end{equation}
the field where the Meissner screening currents \eqref{eq:london} reach $\pm
i_p$ at the edges of the superconductor.

Summarizing, at small fields below $H_s$ the film remains in the Meissner
state (i.e., $n(x) = 0$) with the linear current profile $i(x)$ given by the
London response \eqref{eq:london}. Increasing the field  beyond $H_s$,
vortices enter the film and get pinned, with the corresponding current- and
vortex-density profiles $i(x)$ and $n(x)$ shown in Fig.\
\ref{fig:crit_state}(c); note that $B \approx H$ and $n(x)$ do not necessarily
coincide any more.  The above results will also be relevant in the calculation
of the magnetic moment $m(H)$ in Sec.\ \ref{sec:magnetic} below, where the
hysteretic behavior at decreasing $H$ after a field reversal at $H_0$
will be considered as well.

\subsection{Critical bias current $I = I_c(H)$}\label{sec:critical_current}
We move on to study the critical state preceeding the onset of dissipation,
i.e., when a critical current $I=I_c(H)$ is applied to the superconductor.
While the $I=0$ -- finite $H$ scenario is relevant to magnetization
experiments, the critical state with $I=I_c(H)$ is of great importance for
transport experiments.

In the presence of a finite and homogeneous across $x$ transport current $I =
\int \mathrm{d}x\, i(x)$ fed to the sample, the solution of the Maxwell-London
equation \eqref{eq:maxwell_london} in the Meissner state picks up an
additional term $I/W$,
\begin{equation}\label{eq:current_density}
   i(x) \approx \frac{I}{W} - \frac{cd\,H}{4\pi\lambda^2}x,
\end{equation}
see Figs.\ \ref{fig:current_density}(b) and (c) where the flat and tilted
current distributions at vanishing and finite fields are shown.  Note that the
linear term $\propto -Hx$ induced by the external field $H$ does not contribute
to the transport current.

Criticality is reached when the magnitude of the current density $i(x)$ at
either (or both) edges $x = \pm W/2$ reaches the surface critical current
$i_s$, at which point vortices start penetrating the film.  Using Eq.\
\eqref{eq:current_density}, we find the critical current at vanishing magnetic
field
\begin{equation}\label{eq:I_c_0}
   I_c(0) = I_s \equiv i_s W 
\end{equation}
with vortices/anti-vortices entering from the two edges. At positive
(negative) fields {and positive current bias $I > 0$, the screening term in
Eq.\ \eqref{eq:current_density} favours the nucleation of vortices from the
left (right) edge of the superconductor when
$i(\pm W/2) = I_c/W \mp (cd\,H/4\pi\lambda^2)W/2 = i_s$, and
the critical current reads \cite{Maksimova_1998,Plourde_2001} (we make
repeatedly use of Eqs.\ \eqref{eq:H_s} and \eqref{eq:H_p} to express $cd
W^2/8\pi\lambda^2$ through $i_s/H_s$ or $i_p/H_p$)
\begin{equation}\label{eq:I_c_low_fields}
   I_c(H) = I_s \left(1 -\frac{|H|}{H_s}\right).
\end{equation}
Equation \eqref{eq:I_c_low_fields} defines the critical current in the
Meissner state; its linear cusp is the hallmark of surface pinning and
corresponds to the linear portions of the black solid line in Fig.\
\ref{fig:phase_diagram}. The result is valid as long as the current density
$i(x)$ is larger than depinning everywhere, $i(x) > i_p$, allowing vortices to
traverse the superconductor without stopping.

The field $H \equiv H^\ast > 0$ where $i(x)$ first touches $i_p$ then defines
a new regime where vortices that enter from the left edge stop because of bulk
pinning just before exiting at the right edge. This brings the superconductor
to the onset of the mixed state, with the corresponding critical state shown
by the solid black line in Fig.\ \ref{fig:current_density}(c).  Imposing the
conditions $i(-W/2) = i_s$ and $i(W/2) = i_p$ valid at $H = H^\ast$ on Eq.\
\eqref{eq:current_density}, we obtain the relation $i_s - i_p =
(cd\,H^\ast/4\pi\lambda^2)(W/2)$ that sets the transition between Meissner and
mixed critical states at,
\begin{equation}\label{eq:H_ast}
   H^\ast \equiv \frac{H_s - H_p}{2}.
\end{equation}

Increasing the field beyond $H^\ast$, the critical current-density profile
$i(x)$ turns steeper on the left and the mixed state region expands from the
right edge into the film, see the black solid lines in Figs.\
\ref{fig:current_density}(d) and (e).  Currents $i_p \leq i \leq i_s$ in the
linear region push the vortices into the superconductor until they stop and
accumulate to form a vortex box with constant density $n(x) = H/\Phi_0$ within
the region $[x_0(H),W/2]$. The Maxwell-London equation
\eqref{eq:maxwell_london} then tells that the current density $i(x)$ is
constant within the vortex box, assuming the value $i(x) = i_p$ at
criticality. Matching the linear and constant portions of $i(x)$, we find the
result (see also Appendix \ref{app:current_density} and Eq.\
\eqref{eq:x_m_x_M})
\begin{equation}\label{eq:x_M}
   x_0(H)= -\frac{W}{2} \left(1-\frac{2H^\ast}{H}\right).
\end{equation}
Integrating the resulting current profile $i(x)$, we find the critical current
curve $I_c(H)$ in the form (we define $I_p = i_p W$ and repeat the low-field
result Eq.\ \eqref{eq:I_c_low_fields}; the extension to negative fields $H <
-H^\ast$ is trivial)
\begin{eqnarray}\label{eq:I_c_meissner_state}
   I_c(|H|< H^\ast) &= \displaystyle{I_s \left(1-\frac{|H|}{H_s}\right),}
   \\ \label{eq:I_c_mixed_state}
   I_c(|H| > H^\ast) &= \displaystyle{I_p \left(1 + \frac{(H^\ast)^2}{|H| H_p}\right).}
\end{eqnarray}
The presence of the finite vortex box at fields $|H| > H^\ast$ contributes
with the current $i_p \left[ W/2 - x_0(H) \right]$ to $I_c(H)$, that results
in the nonlinear dependence of the critical current $I_c$ on the field $H$.
In the limit of large magnetic fields $H\gg H^\ast$, $x_0 \to -W/2$ and
$I_c(H)$ saturates to $i_p W = I_p$.  In the absence of bulk pinning, the
critical current drops towards zero as $I_c(H) = I_s\, (H_s/4H)$, see Refs.\
\onlinecite{Maksimova_1998,Plourde_2001}.  Conversely, as bulk pinning becomes
strong, $i_p \to i_s$, the linear peak Eq.\ \eqref{eq:I_c_meissner_state} at small
fields, that is due to surface pinning, gets absorbed into the flat background
due to bulk pinning.  The results \eqref{eq:I_c_meissner_state} and
\eqref{eq:I_c_mixed_state} compactly illustrate the relevance of surface and
bulk pinning: the surface barrier dominates the low-field regime at $|H| <
H^\ast$ and produces a linear cusp in $I_c(H)$, while bulk pinning shapes the
large-field region  $|H| > H^\ast$ and produces a saturation in the tails.

When comparing the above theoretical results for $I_c(H)$ to experiments, we
note that the magnitude of $I_c(H)$ depends, via $I_s$, on the value $i_s$ of
the critical current density at the edges. The latter can vary from sample to
sample due to edge imperfections, rendering the task of predicting the value
of the critical current not straightforward.  This is not the case, however,
for the slope $\mathrm{d} I_c/\mathrm{d} H$ in the Meissner state: taking the
derivative of \eqref{eq:I_c_low_fields}, we find that
\begin{equation}\label{eq:slope}
   \frac{\mathrm{d} I_c(H)}{\mathrm{d} H} 
   \approx - \frac{c\, d}{8\pi}\frac{W^2}{\lambda^2},
\end{equation}
is independent of $i_s$.  The slope of $I_c(H)$ thus assumes a universal
value, determined only by the London length $\lambda$ and the width $W$ of the
film.

\subsection{Finite bias current $0 < I < I_c(H)$}\label{sec:with-transport}

We proceed with the study of the critical state for arbitrary 
bias current $I$ at fixed field $H$.  At small fields $|H| < H^\ast$, the film
always resides in the Meissner phase and thus behaves reversibly, with $n(x) =
0$ and $i(x)$ given by the shifted linear profile \eqref{eq:current_density}.
At larger fields $|H| > H^\ast$, the presence of vortices gives rise to
hysteretic behavior: we will first study the vortex- and current density
profiles $n(x)$ and $i(x)$ for increasing currents $I$ at fixed $H$ in samples
initially prepared at zero bias current $I = 0$ and focus on currents
decreasing from $I_c(H)$ in a second step.  On increasing $I$ from zero at
intermediate fixed fields, $H^\ast < |H| \leq H_s$, the film first displays
zero and then one vortex box (with edges $x_0\leq x_1$), while it goes from
two boxes (with edges $x_0\leq x_1\leq x_2\leq x_3$) to one box at larger
fields $|H| > H_s$, see the differently colored states in
\ref{fig:current_density}(a). Note that the symmetric boxes with coordinates
$\pm x_i$ and $\pm x_o$ at vanishing currents $I = 0$ are replaced by
asymmetric ones involving coordinates $x_0$ to $x_3$ at finite currents $I$.

Without loss of generality, we will consider positive magnetic fields $H$,
such that vortices shift towards the right of the superconductor when $I>0$.
To benefit the reader, we will defer the detailed derivations and expressions
for the edges of the vortex boxes to Appendix \ref{app:current_density} and
focus here on understanding the irreversible evolution of the current-density
profile as the bias current is changed from $0$ to $I_c(H)$ and reverse.

\subsubsection{Increasing the current from $I=0$}\label{ref:prepared_at_0}
In the absence of bias currents and for fields $|H| \leq H_s$, the
superconductor is in the Meissner state.  Upon increasing $I$, vortices first
penetrate inside the sample when the current density
\eqref{eq:current_density} reaches $i_s$ at either of the two edges, cf.\ the
derivation of \eqref{eq:I_c_low_fields}, i.e., along the line
\begin{equation}\label{eq:I_s_vs_H}
   I_s(H) = I_s\left(1 -\frac{|H|}{H_s}\right),
   \quad \text{for}\quad|H| < H_s.
\end{equation}
This (black dotted) line in Fig.\ \ref{fig:current_density}(a) separates the
Meissner- from the mixed states when starting from $I=0$ with $|H| < H_s$ and
coincides with the critical current \eqref{eq:I_c_low_fields} at low fields
$|H| \leq H^\ast$.  As a result, for increasing bias currents $I$, the
Meissner state encompasses the orange and red-dashed regions in Fig.\
\ref{fig:phase_diagram}, see also the orange region in Fig.\
\ref{fig:current_density}(a).

\begin{figure}
        \centering
        \includegraphics[width = 1.\columnwidth]{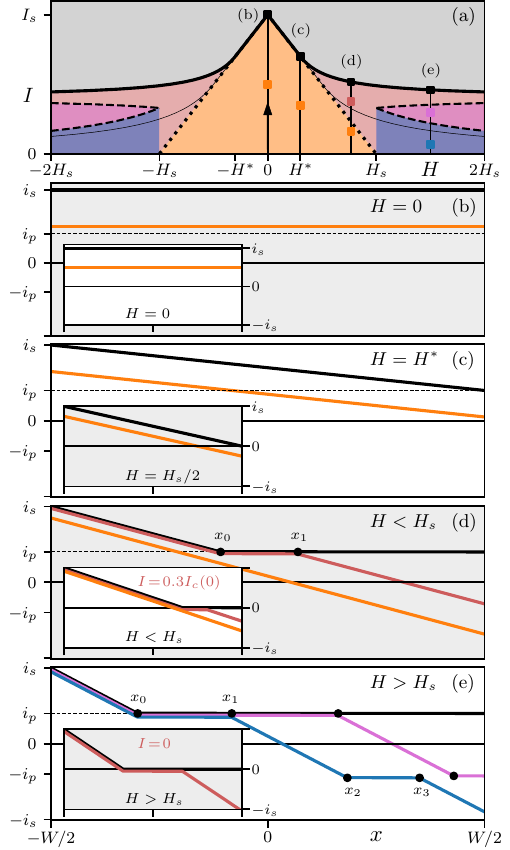}
	\vspace{-10pt} \caption[Current-density profile]{Current density
	profiles at different combinations of $H$ and $I$ (for a film
	initially prepared with $I=0$) corresponding to the colored squares in
	the hysteretic diagram (a), cf.\ Fig.\ \ref{fig:phase_diagram}.  (b)
	At zero field, $i(x) = I/W$ (solid orange line) everywhere and no
	vortex enters the sample until $I = I_c(0)$ (solid black line).  (c)
	For $H \leq H^\ast$, the superconductor remains in the Meissner state
	(solid orange line) until $I = I_c(H)$ (solid black line) is reached.
	At $H = H^\ast$, $i(x)$ reaches $i_s$ $(i_p)$ on the left (right)
	edge; vortices penetrating the sample at $-W/2$ stop just before
	exiting at $+W/2$.  (d) For larger fields $H^\ast < H < H_s$, the
	state of the superconductor depends on the value of the bias current
	$I$.  At small $I<I_s(H)$ (orange), the sample is in the Meissner
	state. For larger currents $I>I_s(H)$ (red), vortices penetrate from
	the left and form a box with $i(x) = i_p$ for
	$x\in\left[x_0,x_1\right]$ (solid black dots).  At the critical
	current $I=I_c(H)$ (black), $x_1$ reaches $W/2$ and vortices start
	flowing across the sample upon further increase of $I$.  (e) For
	fields $H\geq H_s$, the superconductor is always in the mixed state.
	Two symmetric vortex boxes are present at $I=0$.  For $I\leq I_t(H)$
	(blue), the right of the two vortex boxes conserves the vortex number
	attained at $I=0$. At larger currents, vortices start leaving the
	right box (magenta), until this disappears when $I = I_d(H)$; only one
	vortex box remains at currents $I > I_d(H)$ (red). The currents $I_t$
	and  $I_d > I_t$ are shown with black dashed lines in (a).  The insets
	show the current distribution for the same field values (note that
	$\Hast = H_s/2$ at $i_p = 0$) for the case of vanishing bulk pinning,
	i.e., $i_p = 0$.  The black and orange lines refer to transport
	currents $I = I_c$ and $I = I_c/2$ at $i_p = 0$, respectively. }
    \label{fig:current_density}
\end{figure}

For larger bias currents $I > I_s(H)$, vortices accumulate inside the
superconductor, forming a vortex box that extends between $x_0(H)$ and
$x_1(H,I)$; the box boundaries $x_0(H)$ and $x_1(H,I)$ are again obtained by
matching the linear and constant ($i = i_p$) portions of the current profile
derived from the Maxwell-London equation \eqref{eq:maxwell_london} and fixing
the current $I = \int \mathrm{d}x\, i(x)$, see App.\ \ref{app:from_0}.  The
solution $x_0(H)$ is still given by Eq.\ \eqref{eq:x_M}, while $x_1(H,I)$
depends additionally on the current $I$ and reads,
\begin{equation}\label{eq:x_left}
   x_1(H,I > I_s) \!=\! W\left[ \frac{1}{2}-\sqrt{
   \frac{H_p}{H}\frac{I_c(H) - I}{I_p}  } \,\right].
\end{equation}
The first (tiny) vortex box appears on the line $I_s(H)$: for $H = H^\ast$ the
current profile $i(x)$ touches $i_p$ on the right boundary, $i(W/2) = i_p$,
and the first vortex box appears at $x_0 = x_1 = W/2$. With increasing $H$,
the current-density profile $i(x)$ reaches bulk pinning $i_p$ at values $x
<W/2$ and the nucleation of the vortex box moves to the left.  For $H = H_s$,
we have $I_s = 0$ and the current-density profile cuts $\pm i_p$ at two
symmetric points $\pm x = (H_p/H_s)(W/2)$, hence two vortex boxes appear at
these latter points, see Eq.\ \eqref{eq:x_m_x_M}, with $x_i = x_o =
(H_p/H_s)(W/2)$.

The current-density profile for a situation away from $I_s(H)$ is shown in red
in Fig.\ \ref{fig:current_density}(d); the current distribution evolves into
the critical one (black lines) as the bias current is increased to $I_c(H)$
and the vortex box touches the right edge of the sample, allowing vortices to
leave the superconductor.

For fields $|H|\geq H_s$, the superconductor is in the mixed state already for
vanishing bias currents $I = 0$, with two symmetric vortex boxes $\left[x_0 =
-x_o, x_1 = -x_i\right]$ and $\left[x_2 = x_i, x_3 = x_o\right]$ present
inside the sample.  Upon increasing the bias current, the left box $\left[x_0,
x_1 \right]$ expands, as new vortices enter the superconductor from the left
edge, while the right box $\left[x_2, x_3\right]$ shifts towards the right
edge and conserves its size, as vortices therein are trapped by the surface
currents and cannot leave the film.  Again, the boundaries of the vortex boxes
are found by imposing that $i(x)$ (made up from linear and constant portions)
is continuous across the superconductor and that $I\!=\!\int\!\mathrm{d}x\,
i(x)$; the corresponding expressions are given by Eqs.\
\eqref{eq:x_LL}--\eqref{eq:x_RL} and \eqref{eq:x_right} of Appendix
\ref{app:from_0} and a typical current-density profile is shown in blue in
Fig.\ \ref{fig:current_density}(e).

When the bias currents reaches 
\begin{equation}\label{eq:I_t_vs_H}
   I_{t}(H) \equiv I_s \frac{(H^\ast)^2 +  2 H^\ast H_p}{H_s |H|},
\end{equation}
the boundary $x_3$ of the right vortex box touches $W/2$ and the box starts
shrinking as the vortices gradually leave the superconductor.  The
current-density profile typical for this regime is shown in magenta in Fig.\
\ref{fig:current_density}(e) and the corresponding expressions for $x_2,~ x_1$
and $x_0$ are given by Eqs.\ \eqref{eq:x_LR_bis}--\eqref{eq:x_RL_bis} as well
as \eqref{eq:x_right}.  When $x_2 (I,H) = W/2$, the right vortex box has
disappeared completely and only the left box remains; this happens when the
bias current reaches
\begin{equation}\label{eq:I_d_vs_H}
   I_{d}(H) \equiv I_p \frac{(H^\ast)^2 + |H| H_p - H_p^2}{H_p |H|}.
\end{equation}
Equations \eqref{eq:I_t_vs_H} and \eqref{eq:I_d_vs_H} correspond,
respectively, to the lower and upper black dashed curves in the diagram Fig.\
\ref{fig:current_density}(a).  The two curves cross at $H_s$, where i) the
right box touches the sample edge at $W/2$ and ii) its width vanishes.  For
bias currents $I>I_{d}(H)$, the current-density profile displays a single
plateau, i.e., there is only one surviving vortex box whose boundaries $x_0$
and $x_1$ are described by the same expressions \eqref{eq:x_M} and
\eqref{eq:x_left} as obtained for fields $|H|\leq H_s$; finally, the critical
current $I_c(H)$ is reached when $x_1 = W/2$.

\subsubsection{Decreasing the current from $I=I_c(H)$}\label{ref:prepared_at_I_c}

Due to the presence of a finite vortex population inside the superconductor,
the behavior of the mixed state is different when increasing the bias current
$I$ from zero versus decreasing $I$ from $I_c(H)$, that leads to a hysteretic
behavior reminescent of that of the critical state in bulk and flat
superconductors \cite{Tinkham_2004}.  Here, we study the situation where the
transport current $I$ decreases from the critical value $I_c(H)$ at fixed
field $H$ in order to study the hysteretic effects in current $I$, see Sec.\
\ref{sec:magnetic} below for a study of hysteretic effects in $H$. In
preparing the sample, we wish to avoid complications associated with the
dissipative state (e.g., a potential strong rearrangement in the current
profile $i(x)$) and therefore drive the film close to, but below $I_c(H)$
before decreasing the current $I$.

\begin{figure}[t]
        \centering
        \includegraphics[width = 1.\columnwidth]{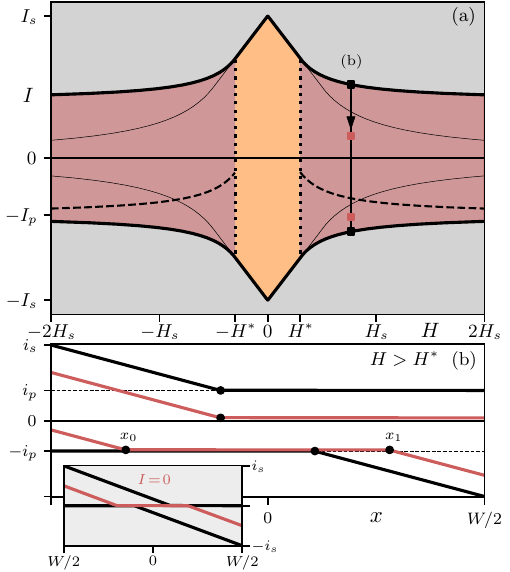}
	\vspace{-10pt} \caption[Current-density profile -- reversed]
	{Current-density profile at different combinations of $H$ and $I$ (for
	a film prepared with $I=I_c(H)$) corresponding to the colored squares
	in the hysteretic diagram (a), cf.\ Fig.\ \ref{fig:phase_diagram}. The
	superconductor remains in the Meissner state (orange) for small fields
	$|H| \leq H^\ast$.  For larger fields $|H| > H^\ast$ the film {\it
	always} resides in the mixed state with one vortex box (red),
	different from the situation with increasing currents from zero
	discussed in Fig.\ \ref{fig:current_density}. (b) At $I = I_c(H)$
	(upper black line) the vortex box extends to the right edge of the
	sample and $i(x > x_0) = i_p$.
	As the current is decreased (upper red line),
	vortices stay trapped and the box remains in place, with the density
	and total number of vortices unchanged.  As a result, the
	current-density profile $i(x)$ is rigidly shifted downwards.  When $I
	= I_c(H) - 2I_p < 0$ (black dashed line in (a)), the plateau in $i(x)$
	reaches $-i_p$, vortices start moving, and the box shifts to the
	left, conserving its size, when the current is further decreased
	(lower red line). 
	At $-I_c(H)$ (lower black line), the current-density
	profile $i(x)$ touches $-i_s$ at the right edge $W/2$ and the vortex
	box has reached the left edge, resulting in an overall inversion
	$i(x) \to -i(-x)$ when going from $I_c(H)$ to $-I_c(H)$. Hence, on
	decreasing the current from $I_c$, the mixed state involves only one
	vortex box that rigidly shifts from the right side to the left side of
	the film.  The inset shows the evolution of the current-density
	profile with decreasing current in the absence of bulk pinning.}
    \label{fig:current_density_rev}
\end{figure}

Starting with small fields $|H| < H^\ast$, the film always remains in the
Meissner state and hence behaves reversibly. Going to higher fields $|H| >
H^\ast$ and preparing the film at $I = I_c(H)$, vortices have populated the
superconductor such that the sample remains in the mixed state for all values
of the bias current $I < I_c(H)$, including the red-dashed region in Fig.\
\ref{fig:phase_diagram}, see also the red region in Fig.\
\ref{fig:current_density_rev}(a).  Decreasing the current $I$ below $I_c(H)$
the vortices that are trapped inside the superconductor remain trapped in the
region $[x_0, W/2]$, with $x_0$ given by Eq.\ \eqref{eq:x_M}, see the upper
solid black curve in Fig.\ \ref{fig:current_density_rev}(b), and the local
density of vortices $n(x)$ does not change as long as $|i(x)|$ remains below
its bulk critical value $i_p$.  As a result, the current-density profile
$i(x)$ is rigidly shifted downwards, see the upper solid red line in Fig.\
\ref{fig:current_density_rev}(b), until the current in the plateau first
reaches $-i_p$ at $I = I_c(H) - 2I_p < 0$ (black dashed line in (a)).  At this
current, the vortices start shifting to the left and a new vortex-free region
with $di/dx \propto -H$ forms close to the right edge of the superconductor
(lower solid red curve in (b)); the position $[x_0(H,I), x_1(H,I)]$ of the
vortex box in the region $I_c - 2I_p > I > - I_c$ is given by Eqs.\
\eqref{eq:x_>_app_I_c}--\eqref{eq:x_<_app_I_c} in Appendix \ref{app:from_I_c}.
Finally, when $I$ is further decreased to $-I_c(H)$, the current density
reaches $-i_s$ at the right boundary $W/2$ in the same instant that $x_0$
touches the left boundary $-W/2$, thus realizing the critical state
appertaining to $-I_c(H)$. Hence, in the process of reducing the current $I$
from $I_c(H)$ to $-I_c(H)$, the vortex box has kept its size and has shifted
from the right side to the left side of the film. Also, note that decreasing
the current from critical, the vortex configuration in the mixed state always
contains a single vortex box, see Fig.\ \ref{fig:current_density_rev}(a),
different from the behavior with increasing currents where the number of
vortex boxes changes between two and one, see Fig.\
\ref{fig:current_density}(a).

\section{Superconducting diode effect in asymmetric 2D thin films}\label{sec:asymmetric_films}

Superconducting films have a long tradition in technological applications,
including Josephson junction-based devices in electronics and many types of
sensors, e.g., for magnetic fields (SQUIDS) or photodetectors (bolometers).
Another application that has been much discussed recently is the
superconducting diode effect (SDE) \cite{Vodolazov_2005, Diez-Merida_2021,
Lyu_2021,Pal_2021, Baumgartner_2022, Bauriedl_2022, Golod_2022, Lin_2022,
Suri_2022, Gutfreund_2023, Margineda_2023, Sundaresh_2023, Hou_2023,
Paolucci_2023, Nakamura_2023, Nadeem_2023, Gaggioli_2023_novo} in 2D
superconductors.  In this type of inductive device, an ac current drive is
applied to produce a dc voltage, the opposite of the usual capacitive setup in
a semiconductor where an ac voltage drive produces a dc current. A simple
mechanism producing this effect is provided by the asymmetric vortex motion
within a superconducting film upon current reversal that produces a
non-reciprocal transport characteristic.  Such a vortex-based SDE can be
realized through asymmetric surface barriers generating different surface
critical current densities $\isL$ and $\isR$ at the left and right edges of
the film, respectively.  This kind of asymmetry is ubiquitous in real thin
films and can result from device imperfections \cite{Plourde_2001} or
deliberate design \cite{Hou_2023}.  Another vortex-based source of
non-reciprocal transport is the ratchet effect in bulk vortex pinning
producing different depinning currents $\ipL$ and $\ipR$ for vortices incident
from the left and right side, respectively.  Such a bulk asymmetry does not
naturally occur in thin film devices, but can be realized in nanostructured
devices \cite{Villegas_2003, SouzaSilva_2006, Gillijns_2007, Lyu_2021,
Lang_2024} and twisted moir\'e superconductors \cite{Gaggioli_2023_novo},
where the degree of asymmetry can directly be controlled by means of an
applied gate voltage. While the surface barrier asymmetry manifests in the low
field peak of $I_c(H)$, the asymmetry in bulk pinning shows up in the tails of
$I_c(H)$ at large fields.

Alternative mechanisms \cite{Nadeem_2023} involving symmetry breaking on a
microscopic level have been discussed in order to generate a superconducting
diode effect. Prominent examples are noncentrosymmetric materials
\cite{Hoshino_2018, Ando_2020, Daido_2022}, finite momentum pairing
\cite{Yuan_2022}, or chiral superconductors \cite{Zinkl_2022}.  At the same
time, use of the SDE has been proposed as a means to infer spontaneous
symmetry breaking in exotic superconductors \cite{Zinkl_2022}. However, care
has to be taken in order not to confuse an intrinsic mechanism with the simple
vortex-based origin where time-reversal- and inversion symmetries are broken
on a macroscopic level, see Ref.\ \onlinecite{Moll_2023} and the discussion
in Sec.\ \ref{sec:sc_diode_exp} below.

In the present section, we provide a simple and quantitative analysis of the
superconducting diode effect in terms of our critical state model, including
also the effect of bulk pinning.  In Sec.\ \ref{sec:asymm_Ic} below, we will
first determine the non-reciprocal critical currents $I_c^+(H)$ and $I_c^-(H)$
for different drives along and opposite to the $y$-axis. Subsequently, we will
find quantitative results for the diode efficiency, see Sec.\
\ref{sec:diode_efficiency}, and then compare our findings with recent
experiments, Sec.\ \ref{sec:sc_diode_exp}.

\subsection{Critical current in asymmetric 2D thin films}\label{sec:asymm_Ic} 

In the following, we will assume that the film asymmetry is such that the
surface critical current density $\isL$ and surface penetration field $\HsL =
(8\pi\,\lambda^2 i_{s,{\scriptscriptstyle L}} /cd W)$, see \eqref{eq:H_s}, at
the left edge are larger than their counterparts $\isR$ and $\HsR$ at the
right edge.  Moreover, for the main part of this section, we will assume that
the surface critical current densities are larger than either depinning
current, i.e., $i_{s,{\scriptscriptstyle (R,L)}} > i_{p,{\scriptscriptstyle
(R,L)}}$.

For such an asymmetric setup, criticality at zero-field $H = 0$ is realized
when the homogeneous current density $i(x)$ in \eqref{eq:current_density}
reaches the smaller critical value $\isR$ at the right boundary $x = W/2$,
hence, $I_c(0) = \isR W$, see the blue line in the inset of Fig.\
\ref{fig:critical_current}. Vortices then enter the sample from the right side
only. Since the application of a field $H > 0$ tilts the current profile, the
current $I$ can be further increased until reaching the critical surface
current $\isL$ at the left film boundary, i.e., $i(-W/2) = \isL$, at a finite
field $\Hmax >0$, see the orange line in the inset of Fig.\
\ref{fig:critical_current}.  Using Eq.\ \eqref{eq:current_density}, the two
conditions $i(-W/2) = \isL$ and $i(W/2) = \isR$ transform to the
relations
\begin{eqnarray}\label{eq:consistency_equation_a}
   \isR W = I_c(0) &= \displaystyle{I_c(\Hmax) - \frac{cd W^2}{8\pi\lambda^2}\Hmax,}\\
   \label{eq:consistency_equation_b}
   \isL W &= \displaystyle{I_c(\Hmax) + \frac{cdW^2}{8\pi\lambda^2}\Hmax.}
\end{eqnarray}
The first equation then relates the peak field $\Hmax$ to the macroscopic
difference in the total currents $I_c(\Hmax) - I_c(0)$,
\begin{equation}\label{eq:H_max}
  \Hmax = \frac{8\pi\lambda^2}{c d W} [I_c(\Hmax) - I_c(0)],
\end{equation}
while the sum and differences of the equations \eqref{eq:consistency_equation_a}
and \eqref{eq:consistency_equation_b}
express the macroscopic quantities $I_c(\Hmax)$ and $\Hmax$ through the
surface parameters $i_{s,{\scriptscriptstyle (R,L)}}$ and
$H_{s,{\scriptscriptstyle (R,L)}}$,
\begin{eqnarray}\label{eq:H_max_micro}
   I_c(\Hmax) &= (I_{s,{\scriptscriptstyle L}} + I_{s,{\scriptscriptstyle R}})/2,\\
   \Hmax &= (\HsL - \HsR)/2.
\end{eqnarray}

In a next step, we account for the presence of bulk pinning by repeating the
strategy adopted for the symmetric film and compute $I_c(H)$.  Defining
$\HastR \equiv (\HsR - \HpR)/2$ and $\HastL \equiv (\HsL - \HpL)/2$ as the
right- and left generalizations of the $H^\ast$ field in Eq.\
\eqref{eq:H_ast}, we find that
\begin{equation}
	I_c(H) \approx \frac{c\,d}{8\pi}\frac{W^2}{\lambda^2} 
        \! \begin{dcases} 
	   \HpR  + {(\HastR)^2}/{|H|} \quad \text{for~} H \leq -\HastR,\\
	   \HsR  + H\quad\text{for~}-\HastR \leq H\leq \Hmax,\\ 
	   \HsL  - H\quad\text{for~} \Hmax \leq H\leq \HastL,\\ 
	   \HpL + {(\HastL)^2}/{H} \quad\text{for~} H \geq \HastL.
		\label{eq:I_c_shifted}
	\end{dcases}
\end{equation}

\begin{figure}
        \centering
        \includegraphics[width = 1.\columnwidth]{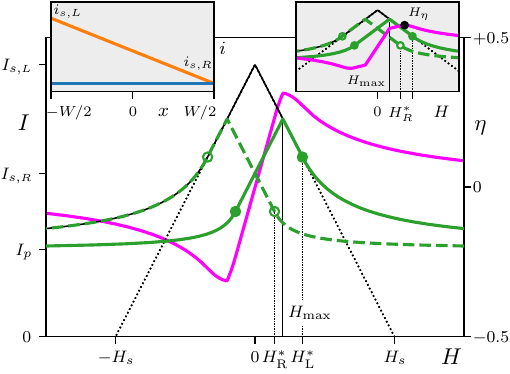}
	\caption[Critical current dependence on field]{Critical current
	$I_c(H)$ as a function of magnetic field $H$ for a symmetric system
	(black line) with $i_p = 2 i_s/5$ and for an asymmetric system  (green
	lines) with $\isL =i_s$, $\isR = 3 i_s/5 $ and $\ipL = \ipR = i_p$.
	In an asymmetric film, the critical current $I_c(H)$ attains its
	maximum at finite field $\Hmax = (\HsL - \HsR)/2$ and $\HastL$ and
	$-\HastR$ delimit the linear segments of $I_c(H)$. For the above
	parameters, $\Hmax = \HsL/5$, $\HastL = 3\HsL/10$, and $\HastR
	= \HsL/10 < \Hmax$.  The critical currents $I^+(H)$ for a
	positive current directed along $y$ (solid green) and the one for a
	negative current $I_c^-(H) = -I_c^+(-H)$ directed along $-y$ are no
	longer the same (the dashed green line shows the modulus
	$|I_c^-(H)|$).  The solid (empty) dots indicate the positive
	(negative) critical currents at the fields $\mp\HastR$ and
	$\pm\HastL$.  The finite peak fields $\pm\Hmax$ produce the
	non-reciprocal transport properties characteristic for a
	superconducting diode; its efficiency (magenta line) is quantified by
	the parameter $\eta$ defined in Eq.\ \eqref{eq:eta_def}. For the
	present choice of parameters, the diode efficiency peaks at $\Hmax$,
	see text. Top left inset: current-density profile $i(x)$ at
	criticality in the asymmetric sample for $H=0$ (blue line) and $H =
	\Hmax$ (orange line). Top right inset: Situation for $\Hmax < \HastR$
	where $\eta$ exhibits a maximum at $H_\eta$ to the right of $\Hmax$.}
    \label{fig:critical_current}
\end{figure}

The overall field dependence of the critical current $I_c(H)$ for an
asymmetric sample with $(\isL - \isR)/(\isL + \isR) = 1/4$ and symmetric bulk
pinning $\ipL = \ipR = i_p = 2 \isL/5$, resulting in a peak field $\Hmax =
\HsL/5$, is shown as a solid green line in Fig.\ \ref{fig:critical_current}.
For a current flow in the opposite direction, the left and right surfaces are
exchanged and the solid green line in Fig.\ \ref{fig:critical_current} is
reflected with respect to the origin, with the peak field $\Hmax$ changing
sign.  Defining $I_c^+(H) = I_c(H)$ and $I_c^-(H)$ as the critical currents
flowing in the positive and negative directions, respectively, we find them
related through $I_c^+(H) =-I_c^-(-H)$, as required by time reversal symmetry.
The absolute value $|I_c^-(H)|$, the dashed green line in Fig.\
\ref{fig:critical_current}, is thus a mirror copy of $I_c^+(-H)$ reflected
with respect to the $H=0$ axis. Asymmetric bulk pinning with
$i_{p,{\scriptscriptstyle R}} \neq i_{p,{\scriptscriptstyle L}}$ results in
different asymptotic currents $I_c (H \gg H^\ast_{\scriptscriptstyle L}) \to
i_{p,{\scriptscriptstyle L}} W$ and $I_c (H \ll -H^\ast_{\scriptscriptstyle
R}) \to i_{p,{\scriptscriptstyle R}} W$.  Hence, while surface asymmetry
manifests in the shift of the peak at small fields, an asymmmetry in bulk
pinning manifests in the tails at large fields.

\subsection{Efficiency of superconducting diode}\label{sec:diode_efficiency}

Given the difference in the positive and negative critical currents,
asymmetric 2D superconducting films display non-reciprocal transport
properties that can be exploited in the fabrication of a superconducting diode
\cite{Vodolazov_2005}.  The non-reciprocal character of these devices is
commonly quantified by the (antisymmetric) diode efficiency
\begin{equation}\label{eq:eta_def}
    \eta(H) \equiv \frac{I_c^+(H) - |I_c^-(H)|}{I_c^+(H) +|I_c^-(H)|}
    = \frac{I_c(H) - I_c(-H)}{I_c(H) + I_c(-H)}
\end{equation}
shown as the magenta line in Fig.\ \ref{fig:critical_current}; in Eq.\
\eqref{eq:eta_def}, we have used that $I_c(H) = I_c^+(H) = -I_c^-(-H)$ and
$I_c(H)$ is given by Eq.\ \eqref{eq:I_c_shifted}.  As the system obeys
time-reversal symmetry, the diode efficiency vanishes for $H=0$.  Moreover, it
saturates to the constant value
\begin{equation}\label{eq:eta_large_field}
    \eta(H \gg H^\ast_{\scriptscriptstyle L}) \approx \frac{\HpL - \HpR}{\HpL + \HpR}
\end{equation}
at large fields where the asymmetric bulk pinning dominates over the effects
of surface barriers and reduces to zero for symmetric bulk pinning $\HpR =
\HpL$, see Fig.\ \ref{fig:critical_current}. The sign of $\eta$ indicates the
polarity of the diode-like behavior.

The full field dependence $\eta(H)$ is straightforwardly obtained by the
expression \eqref{eq:I_c_shifted} for the critical current $I_c(H)$ into the
definition \eqref{eq:eta_def} of $\eta$, see the magenta line in Fig.\
\ref{fig:critical_current}. Below, we focus on the position of the maximum
$\eta_{\max}$ in the efficiency $\eta(H)$ for the case of symmetric bulk
pinning $\HpL = \HpR = H_p$. We start by analyzing the situation at small
asymmetry where $\Hmax < \HastR$, shown in the top-right inset of Fig.\
\ref{fig:critical_current}. In this situation, the cusp of $I_c(H)$ at $\Hmax$
resides within the linear segment of $|I_c^-(H)| = I_c(-H)$ that extends up to
$\HastR$. Similarly, $I_c^+(H) = I_c(H)$ drops linearly up to $\HastL$. As a
result, the numerator $I_c(H) - I_c(-H)$ in Eq.\ \eqref{eq:eta_def} stays
constant in the interval $[\Hmax,\HastR]$ while the denominator decreases,
hence, $\eta(H)$ rises with increasing $H > \Hmax$.  With a continuous
derivative of $I_c^\pm$ across $\HastR$, the efficiency grows further and
reaches its maximum at a field $H_\eta > \HastR$ where
$\partial_{\scriptscriptstyle H} \eta(H_\eta) = 0$, resulting in
\begin{equation}\label{eq:H_eta}
   H_\eta = \HastR\left(\sqrt{\left(\frac{\HastR}{H_p}\right)^2
   + \frac{\HsL}{H_p} } - \frac{\HastR}{H_p}\right) > \HastR.
\end{equation}
The maximal diode efficiency is then found by evaluating the expression
\eqref{eq:eta_def} for $\eta$ at this field,
\begin{equation}\label{eq:eta_max_small_asymm}
    \eta_{\max}= \frac{\left(\HsL^2 -H_p^2\right)/(\HastR)^2
   - 4 \sqrt{1+H_p\,\HsL/(\HastR)^2}}{(H_p + \HsL)^2/(\HastR)^2 + 4}.
\end{equation}
When $\Hmax$ grows larger, the non-analyticity in $I_c(H)$ at $\Hmax$
approaches $H_\eta$ and the above derivation breaks down at $\Hmax =
H_\eta$. For $\Hmax > H_\eta$, the peak in $I_c(H)$ pins the maximum of $\eta$
and hence $\eta_{\max}$ is achieved exactly at $\Hmax$; we then enter the
domain of large asymmetry (shown in the main part of Fig.\
\ref{fig:critical_current}) where
\begin{equation} \label{eq:eta_max_large_asymm}
   \eta_{\max} = \eta(\Hmax) =
   \frac{\HsL \!-\! \Hmax \!-\! H_p \!-\! (\HastR)^2/\Hmax}
        {\HsL \!-\! \Hmax \!+\! H_p \!+\! (\HastR)^2/\Hmax}.
\end{equation}
Let us consider how much the diode efficiency can be increased in a strongly
asymmetric device (but with symmetric bulk pinning $\HpL = \HpR = H_p$) by
decreasing the strength of the weaker (in our convention, the right) surface
barrier $\HsR$ until eventually reaching $H_p = \HsR$.  When this happens, the right
edge does no longer play any role as nucleated vortices are stuck until they
can overcome vortex pinning in the bulk.  The maximal possible diode
efficiency is therefore obtained by replacing the right surface barrier with
the bulk pinning parameter, i.e., by substituting $\HsR = H_p$ in Eq.\
\eqref{eq:eta_max_large_asymm}, yielding
\begin{equation}\label{eq:eta_max_possible}
   \max\left[\eta_{\max}\right] = \eta(\Hmax)\Big|_{\HsR = H_p} 
   =\frac{\HsL - H_p}{\HsL + 3H_p}
\end{equation}
as the maximal possible $\eta$ in a device with finite (symmetric) bulk pinning.

\subsection{Comparison with experiments}\label{sec:sc_diode_exp}

The superconducting diode can be engineered in 2D thin films with asymmetric
surface barriers as proposed and studied in Ref.\ \cite{Vodolazov_2005}. This
type of inductive diode has recently been realized in conventional films (such
as niobium and vanadium) and its functionality has been understood
\cite{Hou_2023}.  Here, we will analyze and discuss two recent experimental
works \cite{Hou_2023,Bauriedl_2022}; in both of these studies, the data for
the critical current $I_c(H)$ look very similar, but the conclusions drawn by
the authors turn out very different. In a first step, we compare the results of our
2D critical state model with their experimental data on the critical current
$I_c(H)$ and subsequently discuss the observed efficiency $\eta$ in the light
of our findings above. The latter has been used to conjecture
\cite{Bauriedl_2022} that the diode effect is not vortex based, as the maximum
in $\eta$ was found at a field beyond $\Hmax$; as demonstrated above, this
finding is nevertheless consistent with our 2D critical-state based vortex
model.

We begin with the analysis of $I_c(H)$ as depicted in Fig.\ 1 of Ref.\
\onlinecite{Bauriedl_2022}; the observed $I_c(H)$ curve shows the typical
shape expected for a surface dominated sample with a sharp cusp at small
fields that compares well with our results in Fig.\
\ref{fig:critical_current}. The experiment was performed in a
$\mathrm{NbSe}_2$ double layer (thickness $d \approx 1.5~\mathrm{nm}$ and
width $W \approx 250 ~\mathrm{nm}$ of the constriction) and clear signatures
of non-reciprocal transport and a superconducting diode effect was observed.
Transforming the maximal critical current $I_c(\Hmax)$ to a current density
$j_c$, we find a value $j_c \approx 3 \cdot 10^6~\mathrm{A/cm}^2$. This has to
be compared with the depairing current density $j_0$ that can be obtained from
phenomenological parameters $\xi$ and $\lambda$ for $\mathrm{NbSe}_2$; while
the bulk value for $\xi \approx 10~\mathrm{nm}$ seems to provide a good
estimate (as $T_c$ and $H_{c2}$ in the film are similar to the bulk
\cite{Gaggioli_2023_novo}), the bulk value $\lambda \approx 250~\mathrm{nm}$
may have to be corrected for the film.  Taking nevertheless these values, we
find a value $j_0 \approx 1.6\cdot 10^7~\mathrm{A/cm}^2$; a reduction $j_s <
j_0$, here by a factor 5, is expected due to surface imperfections and we
consider the agreement quite satisfactory, also with view to our uncertainty
in the size of $\lambda$.

Next, we evaluate the slope $\partial_H I_c(H)$ near the current peak and find
a value $\partial_H I_c(H) \approx 10^{-4}~\mathrm{A/T}$. This has to be
compared with the prediction in Eq.\ \eqref{eq:slope} that includes the
parameters $\lambda_\perp = 2\lambda^2 /d$ and $W$. Using the above parameters
for $d$ and $\lambda$, we find that $\lambda_\perp \approx 80~\mu\mathrm{m}$
that results in a small parameter $W/\lambda_\perp \approx 3 \cdot 10^{-3}$,
quite small indeed.  On the other hand, we can extract $W$ by comparing the
experimental result for the slope $\partial_H I_c(H)$ with the prediction in
Eq.\ \eqref{eq:slope} and find the value $W \approx 100~\mathrm{nm}$, an
outcome that is smaller but still acceptable in comparison with the geometric
value $W \approx 250~\mathrm{nm}$. A similar analysis can be carried out for
the data reported in Ref.\ \onlinecite{Hou_2023} for their vanadium-based
device.

Next, we analyise the experimental data reported for the efficiency.  In Ref.\
\onlinecite{Hou_2023}, Moodera and collaborators fabricated thin film
superconductors with one edge purposefully patterned in order to achieve a
strong diode effect.  For their vanadium device (Fig.\ 1 in Ref.\
\onlinecite{Hou_2023}) they found that $\Hmax \approx 2.8~ \text{Oe}$, while
from the field dependence of the critical current, we estimate that $\HastR
\approx 2.5~ \text{Oe}$ (we identify the weaker edge on the right).
Correspondingly, they report that the diode efficiency is maximal for $H
\approx \Hmax$, see their magenta line in Fig.\ 1(c).  This is in agreement
with our theoretical results for a diode device where the maximum in $\eta$ is
pinned to the cusp in $I_c(H)$ at $\Hmax$.

In Ref.\ \onlinecite{Bauriedl_2022}, the edges of the $\mathrm{NbSe}_2$ thin
film have not been patterned. They find that $\Hmax \approx 10~\text{mT}$.
This is appreciably smaller than the field $\HastR \approx 25~\text{mT}$ that
we estimate from their Fig.\ 1(f).  In addition, they reported a maximal diode
efficiency at a magnetic field $H_\eta \approx 35~\text{mT}$, larger than both
$\Hmax$ and $\HastR$, in agreement with our theory for devices with $\Hmax <
\HastR < H_\eta$.  Based on this finding, i.e., different peak fields for
$I_c(H)$ and $\eta(H)$, the authors concluded that this would rule out a
vortex-based mechanism for the superconducting diode effect. Our analysis
tells instead, that this finding is perfectly consistent with our 2D critical
state model that explains the diode behavior in terms of weakly asymmetric
surface barriers.

The analysis that led to Eqs.\ \eqref{eq:eta_max_small_asymm},
\eqref{eq:eta_max_large_asymm}, and \eqref{eq:eta_max_possible} was carried
out under the assumption of symmetric bulk pinning. Dropping this condition
and assuming asymmetric bulk pinning with $\HpL \neq \HpR$, a finite diode
efficiency can be realized at large fields, see Eq.\
\eqref{eq:eta_large_field}.  Furthermore, it turns out that for a sufficiently
large bulk asymmetry, the maximum $\eta_{\max}$ can be realized in the tails
at large fields rather than at $H_\eta$ or $\Hmax$.  Moreover, a bulk ratchet
effect can give rise to interesting sign reversals of the diode efficiency
\cite{Gillijns_2007, He_2019, Ideue_2020, Jiang_2021, Margineda_2023}.  Within
our 2D critical state model, such a sign change is naturally explained in
terms of opposite asymmetries of surface and bulk pinning, i.e., $\HsL >\HsR$
but $\HpR > \HpL$. Indeed, since low- and large field behaviors are dominated
by surface and bulk, respectively, their change in asymmetry produces a sign
change in $\eta$ as the field $H$ increases.

\section{Magnetic properties of the critical state}\label{sec:magnetic}
We now turn our focus to the magnetic response of 2D thin films with $d \ll
\lambda$, i.e., we determine the magnetic moment $m(H)$ and study the
hysteretic trace for a full sweep $H = 0 \to H_0 \to -H_0~(\to
H_0)$ of the magnetic field along the $I = 0$ axis in Fig.\
\ref{fig:phase_diagram}.  The magnetic moment is generated from the currents
flowing inside the superconductor and is thus related to the self-field
$\mathbf{B}_\mathrm{self}$ discussed in Eq.\ \eqref{eq:maxwell_london_full}.
As previously announced in Sec.\ \ref{sec:maxwell_london}, magnetization
effects are small but nonetheless measurable with sufficiently sensitive
techniques, see the discussion in Sec.\ \ref{sec:film_conclusions} below.
Moreover, our study of the magnetic response will reveal interesting vortex
configurations, such as coexisting vortex- and anti-vortex domains, that
should be accessible to magnetic imaging techniques such as, e.g., the
SQUID-on-tip \cite{Embon_2015}.

Standard magnetostatics tells that the magnetic moment $\mathbf{m}$ of a current
distribution $\mathbf{j}$ is given by the integral
\begin{equation}\label{eq:magnetization_def}
   \mathbf{m} = \frac{1}{2c}\int \mathrm{d}^3\mathbf{r} \,
   \left[\mathbf{r}\times\mathbf{j}(\mathbf{r})\right],
\end{equation}
or equivalently, after integration by parts,
\begin{equation}\label{eq:B-H}
   \mathbf{m} = \frac{1}{4\pi}\int \mathrm{d}^3\mathbf{r} \,
   \left[\mathbf{B}(\mathbf{r}) - \mathbf{H}\right],
\end{equation}
where $\mathbf{B}$ is the induction in the film and $\mathbf{H}$ the external
field.  While our approximation $\mathbf{B} \approx  \mathbf{H}$ was
appropriate for the calculation of the current-density profile $i(x)$, here,
we have to consider the small difference between $B$ and $H$ as produced by
the current-density profile $i(x)$. Although a magnetization experiment is
usually carried out with a squarish sample $W \times W$, here, we keep our 1D
geometry $W  \times L$ with $L \gg W$ of an elongated film in a perpendicular
field to avoid going to a 2D description of the sheet current density
$\mathbf{i}(\mathbf{R})$. The latter flows along $y$ and we find the magnetic
moment $m$ directed along $z$ from \eqref{eq:magnetization_def} to read
\begin{equation}\label{eq:magnetization_def_film}
   m = \frac{L}{c}\int_{-W/2}^{W/2} \!\!\!\!\! \mathrm{d}x\, x \,i(x) 
   = \frac{dL}{4\pi}\int_{-W/2}^{W/2}\!\!\!\!\! \mathrm{d}x\, \left[B(x) - H\right],
\end{equation}
where we have included a factor 2 in order to account for the contribution
from currents parallel to $\pm x$ at the far edges $y = \pm L/2$ of the
sample. Indeed, this last contribution is identical to the contribution from
currents parallel to $\pm y$ and thus contributes a factor two, in agreement
with the result \eqref{eq:B-H} for the total magnetic moment
\cite{Brandt_1993}.

Accounting for the presence of vortices within the type-II film leads to a
hysteretic behavior of the magnetic moment $m(H)$.  Below, we will first
consider the magnetic moment of a pristine (vortex-free) sample in a magnetic
field that is ramped up from zero to $H_0 > H_s$, entering the mixed state at
the penetration field $H_s$.  In a second step, we will reverse the field at a
value $H_0$ with vortices trapped within the film and analyze the completion
of a full loop in the magnetic moment $m(H)$.  We will restrict our analysis
to the case of a symmetric thin film with identical surface barriers and
symmetric bulk pinning.

\subsection{Magnetic moment: increasing fields}\label{sec:m_increasing_fields}

We start with ramping up the external field from zero, as typical in a
zero-field cooled (ZFC) experiment.  For fields smaller than the surface
penetration field $H_s$, the thin film is in the Meissner state and $i(x)$ is
given by Eq.\ \eqref{eq:london}.  Inserting the result into the expression 
\eqref{eq:magnetization_def_film}, we obtain the magnetic moment in the
absence of vortices,
\begin{equation}\label{eq:m_mom_meissner}
   m(|H| < H_s) = - \frac{dWL}{48 \pi}\frac{W^2}{\lambda^2} H.
\end{equation}
The induction $B$ produced by the current density $i(x)$ extends a distance
of scale $W$ in both the $x$ and $z$ directions and hence resembles the response of a
cylinder of diameter $W$ and length $L$. Such a cylinder produces the Meissner
response $m_\mathrm{cyl} = -W^2 L H/8$ (including a demagnetization factor $n
= 1/2$ for a cylinder in a perpendicular field $H$) and expressing the result
\eqref{eq:m_mom_meissner} in terms of the $m_\mathrm{cyl}$, we find that
\begin{equation}\label{eq:m_mom_meissner_zyl}
   m(|H| < H_s) 
	= \frac{W}{3 \pi \lambda_\perp} m_\mathrm{cyl}.
\end{equation}
One should appreciate that the result \eqref{eq:m_mom_meissner_zyl} is quite
large, owing to the fact that we deal with a film in a {\it perpendicular}
field.  Indeed, the response of the film of dimension $W \times W$ produces a
similar moment as that of a cube with dimensions $W$ and there is no geometric
reduction. This is different for a film in {\it parallel} field, where the
moment is down by the geometric factor $d/W$. The only reduction in
\eqref{eq:m_mom_meissner_zyl} is by the factor $W/\lambda_\perp \ll 1$ that
originates from the reduced screening $\lambda \to \lambda_\perp$ due to the
small thickness $d \ll \lambda$ in combination with a restricted film width $W
\ll \lambda_\perp$. For the experiments discussed in Sec.\
\ref{sec:sc_diode_exp} above, sample sizes are about $20~\mu$m and $10~\mu$m
in Refs. \onlinecite{Bauriedl_2022} and \onlinecite{Hou_2023}, respectively,
while $\lambda_\perp \approx 80~\mu$m and $5~\mu$m, hence the ratio
$W/\lambda_\perp$ is of order 0.1 -- 1.

For fields larger than $H_s$, two symmetric vortex-boxes $|x| \in [x_i,x_o]$
as described in Eq.\ \eqref{eq:x_m_x_M} form within the superconductor, see
Fig.\ \ref{fig:m_current_density}, giving rise to the current-density profile
shown in Fig.\ \ref{fig:crit_state}(c).  Inserting the (anti-symmetric)
current-density profile
\begin{equation}\label{eq:iy_H1}
   i(x) = -\frac{cdH}{4\pi\lambda^2} \!
             \begin{dcases} x, ~~ 0 < x \leq x_i(H),\\
                            x_i(H), ~~ x_i(H)\leq x \leq x_o(H),\\
                            x_i(H) + [x\! - \! x_o(H)], 
                             ~~x_o(H) < x,
         \end{dcases}
\end{equation}
into the expression Eq.\ \eqref{eq:magnetization_def_film} results in
\begin{multline}\label{eq:xi_int}
   m(H)= - \frac{dL}{2 \pi \lambda^2} H
   \Biggl[                      \int\limits_0^{x_i}\! \mathrm{d}x\, x^2 
   + x_i(H)                   \!\!   \int\limits_{x_i}^{x_o}\! \mathrm{d}x\, x\\
   +\left[x_i(H) -x_o(H)\right] \!\!  \int\limits_{x_o}^{W/2}\! \mathrm{d}x\, x
                              + \!\!\!  \int\limits_{x_o}^{W/2}\! \mathrm{d}x\, x^2 \Biggr],
\end{multline}
we arrive at the magnetic moment of
the mixed state,
\begin{eqnarray}\nonumber
  & m(H>H_s)= \displaystyle{-\frac{dWL}{48\pi}\frac{W^2}{\lambda^2} 
     \Biggl[-\frac{H}{2} + \frac{H}{2} \biggl(1 - \frac{2H^\ast}{H}\biggr)^3} \\
      &\qquad\qquad\quad 
      \displaystyle{-\frac{H}{2}\left(\frac{H_p}{H}\right)^3 +\frac{3}{2}H_s \Biggr]}
     \label{eq:m_mom_mixed} \\ \nonumber
     =&	\displaystyle{-\frac{dWL}{48\pi}\frac{W^2}{\lambda^2} \Biggl[
   \frac{6(H^\ast)^2}{H} - \frac{4(H^\ast)^3+ H_p^3/2}{H^2} + \frac{3}{2}H_p \Biggr]},
\end{eqnarray}
and antisymmetric in $H$.  The (negative) magnetic moment $m(H)$ as a function
of increasing magnetic field $H$ is shown as a solid black line in Fig.\
\ref{fig:m_moment} for $H_p = H_s/2$.  For moderate bulk pinning strengths
$H_p < H_s/\sqrt{3}$, see below, the magnetic moment $m$ depends on $H$
non-monotonically: at small fields $|H| \leq H_s$, $-m(H)$ is dominated by the
surface barrier contribution and displays a maximum at the onset $H_s$ of the
mixed state with
\begin{equation}\label{eq:m_moment_maximum}
   m_s \equiv m(H_s)= - \frac{dWL}{48\pi}\frac{W^2}{\lambda^2} H_s.
\end{equation}
At large fields $H\gg H_s$, the bulk pinning contribution takes over as
$m(H)$ approaches the asymptotic value
\begin{equation}\label{eq:asymptotic_mag_mom}
   m(H\to \infty)= - \frac{dWL}{32\pi}\frac{W^2}{\lambda^2} H_p.
\end{equation}

As shown in Fig.\ \ref{fig:m_moment}, the magnetic moment 
displays a kink corresponding to a finite jump in the derivative
when vortices first penetrate.  This jump in $\partial_H m$ can be quantified
using Eqs.\ \eqref{eq:m_mom_meissner} and \eqref{eq:m_mom_mixed} by comparing
the left and right derivatives of the magnetic moment at $H_s$,
\begin{equation}\label{eq:left_derivative}
   \partial_H m|_{H_s^-} = - \frac{dWL}{48\pi}\frac{W^2}{\lambda^2}
\end{equation}
and 
\begin{equation}\label{eq:right_derivative}
   \partial_H m|_{H_s^+}
   = \frac{dWL}{96 \pi}\frac{W^2}{\lambda^2}\Biggl[1 
   - 3\frac{H_p^2}{H_s^2}\Biggr],
\end{equation}
yielding the jump
\begin{equation}\label{eq:der_jump}
   \partial_H m|_{H_s^+} \!-\! \partial_H m|_{H_s^-}\!
   = \frac{dWL}{32\pi}\frac{W^2}{\lambda^2}\Biggl[1 - \frac{H_p^2}{H_s^2}\Biggr].
\end{equation}
For negligible bulk pinning $H_p\ll H_s$, we find the universal result that
the right derivative \eqref{eq:right_derivative} is minus half the left value
\eqref{eq:left_derivative}.  For larger fields $H_p$, the absolute value of
the right derivative becomes smaller, until it changes sign at
$H_p=H_s/\sqrt{3}$.  For stronger pinning $H_p>H_s/\sqrt{3}$, the magnetic
moment changes monotonically with $H$, with the kink in $m(H)$ disappearing as
$H_p \to H_s$, see the inset of Fig.\ \ref{fig:m_moment}.  For larger pinning
strengths $H_p > H_s$, the surface barrier does not play any role, as the
effective penetration of vortices inside the thin film is determined by the
field $H_p$ at which they start moving.  This situation can thus be reduced to
the case $i_s = i_p$ and is described by Eqs.\ \eqref{eq:m_mom_meissner} and
\eqref{eq:m_mom_mixed} by substituting $H_s$ with $H_p$.

\subsection{Magnetic moment: decreasing fields}\label{sec:m_decreasing_fields}
After increasing the field in a pristine film, we reverse $H$ at the value
$H_0$.  We will mainly focus on the situation where surface pinning is
dominant, i.e., $H_s > H_p$. Reversing the field in the Meissner region $H_0
\leq H_s$, the film responds reversibly. For a reversal field $H_0 >
H_s$, vortices occupy a finite portion of the thin film and do not immediately
leave the sample upon decreasing the magnetic field, see Fig.\
\ref{fig:m_current_density} showing the current density $i(x)$ and vortex
density $n(x)$ profiles in the right and left halves of the film,
respectively. The system shows memory effects with the magnetic moment
deviating from Eq.\ \eqref{eq:m_mom_mixed}, resulting in an hysteretic $m(H)$
loop as shown in Fig.\ \ref{fig:m_moment}.

\begin{figure}
        \centering
        \includegraphics[width = 1.\columnwidth]{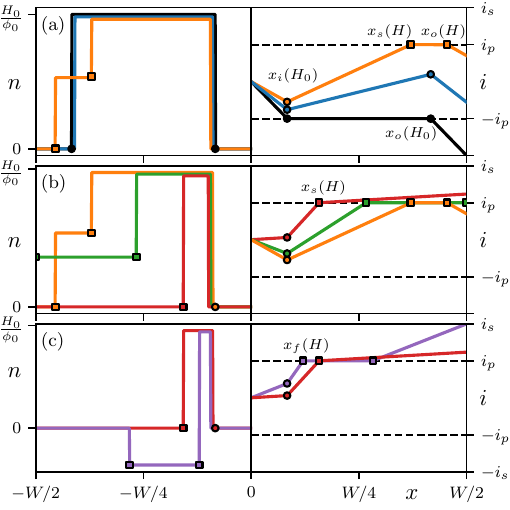}
	 \caption[Vortex- and current-density profiles along the magnetization
	 loop] {Vortex- and current densities $n(x)$ (left) and $i(x)$ (right)
	 across a symmetric 2D thin film with $H_p = H_s/2$ for a magnetic field
	 decreasing after a field reversal at $H_0 = 2(H_s + H_p)$.  The
	 corresponding magnetic moments $m(H)$ are marked in Fig.\
	 \ref{fig:m_moment}, with the colored dots corresponding to the
	 different fields $H$ in the present figure.  The vortex- and current-density
	 profiles at $H_0$ are shown in black. Vortices occupy the region
	 $|x|\in\left[x_i(H_0),x_o(H_0)\right]$ with edges (circles) given by
	 Eq.\ \eqref{eq:x_m_x_M}. (a) As the magnetic field is decreased from
	 $H_0$ (blue lines), vortices initially cannot move and $n(x)$ is
	 unchanged (overlaping density lines are artificially shifted for
	 clarity); the corresponding current density $i(x)$ assumes a positive
	 slope ($\propto -(H-H_0)$, see Eq.\ \eqref{eq:maxwell_london}) within
	 the vortex box.  (b) As $H$ is decreased below $H_1$ (Eq.\
	 \eqref{eq:H_1}, orange lines), the outermost vortices rearrange to
	 form a new region $|x|\in \left[x_s(H), x_o(H)\right]$ (squares) with
	 lower density $n(x) = H/\Phi_0$ such that the total vortex number is
	 conserved; edges (marked by squares) are given by Eqs.\
	 \eqref{eq:y_<} and \eqref{eq:y_>}.  Within this new vortex box the
	 current density is constant and equal to $i_p$. Decreasing $H$ below
	 $H_2$ (Eq.\ \eqref{eq:H_2}, green lines), the outer boundary of the
	 vortex box touches the sample edges and vortices start leaving the
	 sample. (c) For negative fields $H <0$ (red lines), vortices stay
	 trapped only close to the center of the film for
	 $|x|\in\left[x_i(H_0), x_s(H)\right]$.  Anti-vortices start entering
	 the sample when $H\leq H_3$ (Eq.\ \eqref{eq:H_3}, purple lines) and
	 reside next to the original vortices in the region
	 $|x|\in\left[x_f(H), x_o(H)\right]$ (squares).  Anti-vortices are shown with
	 negative vortex density $n(x)$.  Finally, the original vortex domain
	 is completely annihilated and replaced by a corresponding anti-vortex
	 box at $H=-H_0$.
}
    \label{fig:m_current_density}
\end{figure}

The current distribution across the thin film along the irreversible part of
the magnetic loop can be obtained by integrating the Maxwell-London
equation \eqref{eq:maxwell_london} with the appropriate vortex distribution
$n(x)$, see Fig.\ \ref{fig:m_current_density}.  Inserting the resulting
current distribution $i(x)$ into the equation
\eqref{eq:magnetization_def_film} for the magnetic moment $m(H)$ then produces
the curves shown in Fig.\ \ref{fig:m_moment}.  To benefit the reader, we will
leave the lengthy but straightforward derivations of the analytical results
behind these figures to the Appendix \ref{app:m_moments} and focus here on
understanding the irreversible behavior of $n(x)$, $i(x)$, and $m(H)$ from the
underlying interplay of surface barrier- and bulk pinning effects.

For a magnetic field $H$ just below $H_0$, the magnitude of the current
density in the vortex box drops below $i_p$. Due to pinning, the vortex
density $n(x)$ then remains unchanged from the one at $H_0$ and assumes the
value $n_0 = H_0/\Phi_0$ within the vortex boxes $|x| \in
\left[x_i(H_0),x_o(H_0)\right]$, see Eq.\ \eqref{eq:x_m_x_M}, and zero
elsewhere. This is shown on the left part of Fig.\ \ref{fig:m_current_density}
with the vortex density $n(x)$ at the reversal field $H_0$ (black line)
overlapping with the one at $H$ (blue line).  The Maxwell-London equation
\eqref{eq:maxwell_london} then tells that the current $i(x)$ displays a
negative slope $\propto -H$ in the vortex-free region and a positive slope
$\propto -(H-n\Phi_0) = -(H - H_0)$ in the vortex-boxes, see the blue line in
the right part of \ref{fig:m_current_density}; the corresponding vortex- and
current-density profiles are given by Eqs.\
\eqref{eq:ny_H1}--\eqref{eq:iy_H1}.  Remarkably, and differently from the
critical state in the bulk, vortices in this configuration are subject to a
finite current density $i$ of magnitude smaller than $i_p$.  In this range of
magnetic fields, the magnetic moment is given by Eq.\ \eqref{eq:m(H)_abv_H1}
that depends linearly on $H$ with the Meissner slope
\eqref{eq:m_mom_meissner}, see the blue lines in Fig.\ \ref{fig:m_moment}
(with corresponding results at $-H$).

\begin{figure}
        \centering
        \includegraphics[width = 1.\columnwidth]{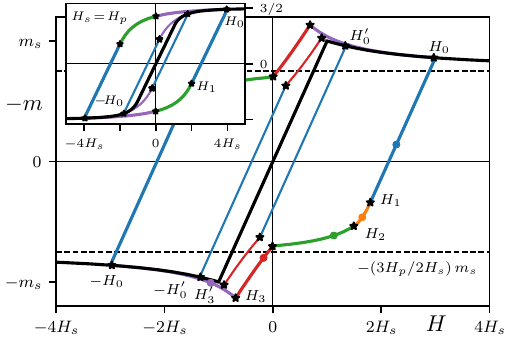}
	 \caption[Magnetic moment of a 2D thin film superconductor]{Negative magnetic
	 moment of a 2D thin-film superconductor with $H_p = H_s/2$ for different
	 reversal fields $H_0$: thick lines for $H_0 = 2(H_s + H_p)$, thin
	 lines for $H_0' = 0.9 (H_s + H_p)$.  The colored dots in the large
	 loop identify the fields corresponding to the vortex- and
	 current-density profiles in Fig.\ \ref{fig:m_current_density}; black
	 stars mark the fields $H = 0$ and $H_{0,1,2,3}$ where vortex boxes
	 undergo changes, see below.  While ramping up the magnetic field
	 (black line) with the sample initially in the pristine state, the
	 magnetic moment grows linearly with $H$ according to Eq.\
	 \eqref{eq:m_mom_meissner} until vortices first penetrate inside the
	 film at $|H|= H_s$ where $m(H)$ displays a kink, see Eq.\
	 \eqref{eq:der_jump}.  The magnetic moment in the mixed state is given
	 by Eq.\ \eqref{eq:m_mom_mixed} and saturates to a constant
	 $(3H_p/2H_s) m(H_s)$ given by bulk pinning at large fields $H$.
	 Decreasing the magnetic field after a field reversal, the system
	 remembers its state at $H_0$ and the magnetic loop exhibits
	 hysteretic behavior. The magnetic moment $m(H)$ goes through
	 different regimes (properly color coded) with vortex- and
	 current-density profiles shown in Fig.\ \ref{fig:m_current_density}.
	 The vortex-density profile changes below $H_1$, first keeping the
	 vortex number conserved (orange), then vortices first leave the
	 system (below $H_2$, green) and then hold to a thinner box close to
	 the center (below $H=0$, red).  Subsequently, anti-vortices enter the
	 sample for negative field values $H\leq H_3 < 0$ and progressively
	 annihilate trapped vortices (purple) until all vortices have been
	 replaced by anti-vortices upon reaching $-H_0$.  Reversing the field
	 at a smaller value $H_s < H_0' < H_s + H_p$ removes the orange and
	 green intermediate regimes of the loop, with blue and red lines
	 joining at $H_1$. The top left inset shows the magnetic moment (in
	 units of $m_s$) at strong bulk pinning $H_p = H_s$.}
    \label{fig:m_moment}
\end{figure}

The above state remains valid as long as vortices do not move away from their
original positions, i.e., as long as the current-density profile $i(x)$ within
the vortex box (that has positive slope) remains in the undercritical domain
with $-i_p < i(x) < i_p$.  As the magnetic field is lowered from $H_0$, this
condition is first broken when $i(x_o(H_0)) = i_p$, i.e., when $H$ takes the
value
\begin{equation}\label{eq:H_1}
   H_1 = \frac{H_0-(H_s + H_p)}{H_0 - (H_s - H_p)} \, H_0
\end{equation}
that determines the transition from the blue to the orange curves in Figs.\
\ref{fig:m_current_density} and \ref{fig:m_moment}.  Below $H_1$, the
vortex-density profile $n(x)$ is changed as the vortices at the outer edge of
the box move under the critical force $i =  i_p$ from their original position
$x \leq x_o(H_0)$ towards the edge. The vortex box thus devides into high- and
low-density parts with densities $n_0 = H_0/\Phi_0$ and $n = H/\Phi_0$,
resulting in a two-stage vortex box with a density step at $x_s(H)$ and
inner/outer edges at $x_i(H_0)$ and at $x_o(H)$.  The position $x_s(H)$
derives from tracing the current profile $i(x)$ from $x=0$ to $x_i(H_0)$ (with
slope $\propto -H$) to the point $x_s(H)$ (with slope $\propto -(H-H_0)$)
where it reaches the critical value $i_p$; the condition $i(x) = i_p$ defines
the location
\begin{equation}\label{eq:y_<}
   x_s(H) = \frac{W}{2}\frac{2H_p}{H_0 - H}
\end{equation}
of the density step.  As long as the resulting two-stage vortex-box
$[x_i(H_0), x_s(H)]$ with density $n_0$ and $[x_s(H), x_o(H)]$ with density
$n$ does not extend to the film boundary, the total number of vortices remains
conserved.  To find the box's outer edge $x_o(H)$, we impose the condition for
vortex number conservation in the form
\begin{multline}\label{eq:n_vortices_cons}
   \frac{H_0}{\Phi_0}(x_o(H_0) - x_i(H_0)) = \frac{H_0}{\Phi_0}(x_s(H) - x_i(H_0))\\
   + \frac{H}{\Phi_0}(x_o(H) - x_s(H)),
\end{multline}
yielding 
\begin{equation}\label{eq:y_>}
   x_o(H) = \frac{W}{2}\frac{H_0 - (H_s + H_p)}{H}.
\end{equation}
The current density inside the region $|x| \in\left[x_s(H),x_o(H)\right]$ is
identical to $\pm i_p$ and assumes a linear slope $\propto -H$ between $\pm
x_o(H)$ and the edges $\pm W/2$ of the thin film.  The corresponding vortex-
and current densities are given by Eqs.\ \eqref{eq:ny_H2}--\eqref{eq:iy_H2}
and are shown in orange in Fig.\ \ref{fig:m_current_density}.  Due to the
conservation of the total vortex number, the areas below the black and orange
curves in the $n(x)$ plot are identical.  The magnetic moment $m(H)$ turns
non-linear in the interval $H_2 \leq H \leq H_1$, see Eq.\
\eqref{eq:m(H)_abv_H2} and the orange segment in Fig.\ \ref{fig:m_moment}.
The boundary $H_2$ of this domain is determined by the condition $x_o(H) = W/2$
where vortices start leaving the film,
\begin{equation}\label{eq:H_2}
   H_2 = H_0-H_s - H_p.
\end{equation}

At $H_2$, we enter the next regime where vortices are free to exit the sample
and their number is no longer conserved.  The corresponding current- and
vortex densities are shown in green in Fig.\ \ref{fig:m_current_density}, with
the density $n(x)$ in the inner (outer) vortex box given by $n_0(x) =
H_0/\Phi_0$ ($n(x) = H/\Phi_0$).  The inner boundary and the location of the
density step in the two-stage box are still given by $x_i(H_0)$ and $x_s(H)$,
while the outer boundary is pinned to the edge, $x_o(H) = W/2$; the vortex- and
current-density profiles are given in Eqs.\ \eqref{eq:ny_0}--\eqref{eq:iy_0}.
The constant currents $\pm i_p$ flowing in a large part of the sample produce
a reduction in the field dependence of the magnetic moment $m(H)$, see Eq.\
\eqref{eq:m(H)_abv_0}, and the corresponding (green) portion of the
magnetic moment $m(H)$ in Fig.\ \ref{fig:m_moment} turns more flat. The orange
segment between $H_1$ and $H_2$ connects smoothly the blue and green segments
at larger and smaller fields, with no change in the derivative of $m(H)$ at
$H_1$ and $H_2$ (but a change in curvature).

Vortices continue leaving the film until $H = 0$; at vanishing field, the
density $n = H/\Phi_0$ in the outer vortex box has dropped to zero and all
vortices at criticality (i.e., where $i =  i_p$) have left the film.  For
negative fields $H < 0$, a thin vortex box with original density $n_0 =
H_0/\Phi_0$  survives in the intervals $|x| \in [x_i(H_0), x_s(H)]$, with
$x_s(H)$ still determined by tracing the current profile starting at $x = 0$
and using the condition $i(x) =  i_p$, what results in the expression
\eqref{eq:y_<} for $x_s$. In the remaining part $|x|\geq x_s(H)$, $i(x)$ changes
linearly with positive slope $\propto -H$, see the red lines in Fig.\
\ref{fig:m_current_density} and Eqs.\ \eqref{eq:ny_H3} and \eqref{eq:iy_H3}
for the vortex- and current-density profiles.  Within the interval $H_3 \leq H
\leq 0$, the magnetic moment $m(H)$ changes close to linearly with field $H$
and is given by Eq.\ \eqref{eq:m(H)_abv_H3}; furthermore, $m(H)$ displays a
kink at $H = 0$ and overshoots the maximal Meissner value $m(H_s)$ given in
Eq.\ \eqref{eq:m_moment_maximum} close to $H_3$.

The last hysteretic regime along the magnetic loop is encountered when
anti-vortices start penetrating from the film edges.  This happens when the
above current profile $i(x)$ (traced from $x = 0$, see red line in Fig.\
\ref{fig:m_current_density}) touches $\pm i_s$ at $\pm W/2$, resulting in the
field
\begin{equation}\label{eq:H_3}
   H_3 = \frac{H_0 \!-\! H_s \!-\! H_p}{2}   \left[1 - 
   \sqrt{1 + \frac{8 H_0 H^\ast}{(H_0 \!-\! H_s \!-\! H_p)^2}}\right].
\end{equation}
The vortex-density profile $n(x)$ again splits into two parts, the leftover
vortices with density $n_0 = H_0/\Phi_0$ in the region $x_i(H_0) \leq |x| \leq
x_f(H)$ and the incoming anti-vortices in the region $|x| \in [x_f(H),
x_o(H)]$. The vortex--anti-vortex front $x_f(H)$ and the boundary $x_o(H)$
again derive from tracing the current profile with the appropriate slopes,
once from $x = 0$ till cutting $\pm i_p$ and the other time from $W/2$ where
$i(\pm W/2) = \pm i_s$ and $i(\pm x_o(H)) = \pm i_p$.  The resulting vortex-
and current densities are given by Eqs.\ \eqref{eq:ny_H4}--\eqref{eq:iy_H4}
and are shown in purple in Fig.\ \ref{fig:m_current_density}.  At $H_3 < 0$,
the magnetic moment curve $m(H)$ in Fig.\ \ref{fig:m_moment} displays another
kink and, for sufficiently weak pinning, the slope of $m(H)$ changes sign at
$H_3$ (see Eqs.\ \eqref{eq:left_derivative} and \eqref{eq:right_derivative}
for a quantitative analysis of the kink at $H_s >0$).  Correspondingly, the
absolute value of $m(H)$ starts decreasing again, see Eq.\
\eqref{eq:m(H)_abv_H4} and the purple curves in Fig.\ \ref{fig:m_moment}.
Note that the magnitude $|H_3| < H_s$, i.e., anti-vortices enter the film at a
lower field amplitude $H_3$ as compared with the field $H_s$ for vortex entry.
Hence, the presence of vortices inside the film lowers the critical entry
field for anti-vortices.

As shown with the purple line in the left panel of Fig.\
\ref{fig:m_current_density}, in this field regime the vortex- and anti-vortex
domains are adjacent to one another at $\pm x_f(H)$. With further decreasing
field $H < H_3$, the vortex box is gradually shrinking through incoming
anti-vortices---vortex--anti-vortex annihilation then moves the boundary
$x_f(H)$ towards the film center until all vortices have disappeared when
reaching $x_i(H_0)$. This happens at $H = -H_0$ where the entire original
vortex box produced at $H_0$ has been replaced by the congruent anti-vortex
box and all memory of the field reversal at $H_0$ is erased.

Above, we have implicitly assumed that vortex--anti-vortex annihilation is
stabilized at $\pm i_p$, i.e., by bulk pinning.  Indeed, the
vortex--anti-vortex attraction quantified by the current density $\sim i_0
\xi/s$ at separation $s$ should be (over-)compensated by the bulk pinning
current $i_p$ at a scale $s \sim a_0$; at very weak bulk pinning, $s \sim i_0
\xi/i_p$ grows beyond $a_0$ and an additional vortex-free region 
separates the vortex box from the anti-vortex box.

So far, we have assumed a reversal field $H_0$ larger than $H_s + H_p$, that
implies positive values for the fields $H_1$ and $H_2$ given by Eqs.\
\eqref{eq:H_1} and \eqref{eq:H_2}. Reversing the field at smaller fields $H_s
< H_0 < H_s + H_p$, we find that $H_1 < 0$ and vortices at $x_s(H)$
immediately leave the film for fields $H \leq H_1$.  Correspondingly, the
intermediate regimes $H_2\leq H \leq H_1$ and $0\leq H \leq H_2$ are not
realized any longer; referring to Figs.\ \ref{fig:m_current_density} and
\ref{fig:m_moment} the behavior of $n(x)$, $i(x)$, and
$m(H)$ jumps directly from 'blue' to 'red' at $H = H_1 < 0$. Note that the
current distribution $i(x)$ at $H < H_1$ differs from the one at fields larger
than $H_1$ only by the substitution $x_o(H_0) \to x_s(H)$ for the outer
boundary of the vortex-box.

To conclude this section, let us relax the condition of a strong surface
barrier and consider the possibility that bulk pinning prevails with $H_p >
H_s$.  This corresponds to the scenario where the vortex entry to the film is
hampered by bulk pinning and the critical current density $i_s$ does not play
any role.  As discussed at the end of Sec.\ \ref{sec:diode_efficiency}, this
regime is successfully described with the help of the results from this
section upon substitution of $H_s$ with $H_p$.  In this limit, the field $H_2$
reduces to $H_1$ for any value of $H_0$, as the original vortex box extends to
the sample edges and vortices start leaving the sample as soon as they can
move from their original positions when the current density reaches the bulk
critical value $\pm i_p$.  In addition, $H_3 = 0$ when $H_0 > 2H_p$, while
$H_3 = H_1$ when $H_1 < 0$, i.e., for $H_0 < 2H_p$, see the inset of Fig.\
\ref{fig:m_moment} where $H_p = H_s$.  In the first case, the outermost
vortices reduce their density as the field is decreased below $H_1$ (the
magnetic moment changes from `blue' to `green' type behavior in Fig.\
\ref{fig:m_moment}, see also Fig.\ \ref{fig:m_current_density}), and are
gradually replaced by anti-vortices as $H$ turns negative (`green' to `purple'
type behavior).  In the second, anti-vortices start penetrating into the
sample as soon as vortices can escape their pinning centers (direct `blue' to
`purple' type behavior).  As for the previously considered scenario with $H_p
< H_s$, vortex- and anti-vortex domains coexist for $-H_0\leq H\leq
\min(0,H_1)$.  Finally, because anti-vortices replace vortices immediately as
these exit the sample at $H=0$ or $H=H_1$, the magnetic trace $m(H)$ displays
no kink upon sweeping the external magnetic field when the surface barrier is
weaker than bulk pinning.

\section{Conclusions}\label{sec:film_conclusions}

We have presented a 2D critical state model that describes the
macrophenomenological properties of 2D thin films with thickness $d \ll
\lambda$. The geometrically suppressed screening introduces the new
characteristic length $\lambda_\perp = 2 \lambda^2/d \gg \lambda$ that easily
exceeds the film's width $W$.  In such films with $d \ll \lambda$ and $W \ll
\lambda_\perp$, the current profile $i(x)$ can be found from a straightforward
integration of the Maxwell-London equation \eqref{eq:maxwell_london} in the
absence of self-field effects \eqref{eq:self_field}.  The driving term in the
Maxwell-London equation \eqref{eq:maxwell_london} depends on the
vortex-density profile $n(x)$ that we find by including both surface barriers
governing vortex entry as well as bulk pinning trapping vortices within the
film. The solution of the Maxwell-London equation then involves only linear
(Meissner state) and constant (mixed state) segments of current density;
together with the vortex profile $n(x)$ (made from vortex boxes), these
segments are self-consistently combined into the current profile $i(x)$. The
results of this 2D critical state model differ pronouncedly from the results
of the Bean model describing bulk samples and are quantitatively different
from the usual ($d > \lambda$) flat sample macrophenomenology that is dominated
by demagnetization- and geometric edge effects.

We have made use of our 2D critical state model to analyze and solve various
tasks, the critical current $I_c(H)$ in both symmetric and asymmetric 2D thin
films, the determination of the (hysteretic in $I$) state at arbitrary
currents $-I_c < I < I_c$ and (fixed) fields $H$, and the hysteretic (in $H$)
magnetic moment $m(H)$.  In the resulting phenomenology, effects of surface
pinning manifest at small fields, specifically a linear cusp in $I_c(H)$, Eq.\
\eqref{eq:I_c_meissner_state}, while bulk pinning manifests in the tails of
$I_c(H)$ at large fields, Eq.\ \eqref{eq:I_c_mixed_state}.  In asymmetric thin
films with edges of different surface pinning strength and/or ratchet type
bulk pinning, the current transport is non-reciprocal, with the peak maximum
in $I_c(H)$ shifted away from zero to $\Hmax$, see \eqref{eq:H_max}.  This
non-reciprocity lends itself to the fabrication of superconducting diodes;
here, we have extended previous analysis \cite{Maksimova_1998,Plourde_2001,
Vodolazov_2005} to include the effect of bulk pinning and have determined the
diode's efficiency. The comparison with recent experiments shows good
agreement between data and theory.

The magnetic moment $m(H)$ of 2D thin films has remained uncharted
territory so far, both with respect to experiment and theory. While the
induced current $i(x)$ in the film is small (and can be neglected in the
self-field contribution \eqref{eq:self_field} to the Maxwell-London equation),
it is this small current that determines the magnetic moment $m(H)$ in
\eqref{eq:magnetization_def_film}. When surface pinning dominates over bulk
pinning, vortex entry into the film manifests as a cusp in the magnetic moment
curve $m(H)$, see Eq.\ \eqref{eq:der_jump}. Upon reversing the field in the
mixed state at $H_0$, the trapped vortex box evolves in several steps, with
field regimes where the box remains invariant, where it splits into two
adjacent boxes, where vortices leave as the box reaches the film's edge and
the box shrinks, where anti-vortices annihilate the remaining shrunk box,
until all the original vortex box at the field reversal $H_0$ has transformed
to a corresponding anti-vortex box at $-H_0$ and all memory of the initial
state has been lost.

In our study, we have made several assumptions that may be overcome in future
work. E.g., we have assumed that the external current $I$ is fed homogeneously
to the film. In a real sample, though, the contact may feed the current to the
film in a non-uniform way, particularly when the contact is small compared to
the film width. Second, we remind that our calculation for the magnetic loop
has focused on symmetric films, that is the exception rather than the rule as
asymmetry in surface pinning occurs quite naturally. Such asymmetry, though,
introduces further complexity into the vortex- and current-density profiles,
that may be worth analyzing once experiments are being carried out.
Furthermore, we have kept the stripe geometry in the calculation of the
magnetic moment for simplicity, although typical samples in this type of
experiments are rather of squarish shape.

An interesting question is whether and how the current profile $i(x)$ and the
magnetic moment $m$ could be measured in an experiment, as these quantities
are small due to the film's small thickness $d$; furthermore, the film
dimensions (in the $\mu$m) range are usually quite small.  Assuming typical
current densities $j \sim 10^6~\mathrm{Acm}^{-2}$ and thicknesses $d$ in the
few nanometer range, we expect local (parallel) fields $B_x \sim (2\pi/c) i_y$
of order a fraction of a Gauss (e.g., $0.3~\mathrm{G}$ for $d =
5~\mathrm{nm}$).  A correspondingly high sensitivity is provided by the
three-junction SQUID on tip (3JSOT) device introduced in Ref.\
\onlinecite{Anahory_2014} that has been developed for simultaneous measurement
of parallel and perpendicular (to the film) field components. Using the 3JSOT
as a scanning probe at a distance of order $10~\mathrm{nm}$ above the film,
one should be able to image the current-density profile $i_y(x) = i(x)$ in
typical samples with widths from $100~\mathrm{nm}$ to $10~\mu$m, given the
reported ac resolution of the device on the order of $0.1~\mathrm{G}$. Using
similar sensitive devices allows to measure the magnetic moment $m$ of the
film, e.g., via integration of the magnetization $M_z(\mathbf{R})$ map
\cite{Grover_2022}.

Another concern is the application of the macrophenomenological theory to
mesoscopic sized samples that appear quite numerously in recent experimental
setups. Considering the relevant field scale $H_s$, see Eq.\
\eqref{eq:penetration_field}, we can extract a relevant vortex distance $a_0
\sim \sqrt{\xi W}$, that produces a small number $W/a_0 \sim \sqrt{W/\xi}$ of
vortex rows. With typical values $\xi \sim 10~\mathrm{nm}$ and $W \sim
\mu\mathrm{m}$, we arrive at order 10 vortex rows and our simple averaging in
a vortex profile $n(x)$ may have to be reconsidered.  Nevertheless, our
results provide a simple and effective first description of the phenomenology
of 2D thin films, and improving upon our macro-modelling may require
numerical tools.

An interesting observation concerns the non-uniform (linear) current profile
$i(x)$ that we found in the 2D thin film, a result deriving from the reduced
screening. This linear profile contrasts to the usual constant current density
profile in the bulk Bean model and may bear interesting consequences when
driving the film beyond critical. Indeed, in the dissipative state, flux-flow
or normal, we expect a uniform current that is driven by a (uniform) electric
field. The transition from the superconducting to the dissipative state in the
film then is associated with a sudden rearrangement of the current profile
that is expected to give rise to voltage jumps and hysteretic effects.

\section*{Acknowledgments}
We thank Mikhail Feigelman, Alex Gurevich, and Jagadeesh Moodera for inspiring
discussions. We acknowledge financial support of the Swiss National Science
Foundation, Division II and the support of the EU Cost Action
CA16218 (NANOCOHYBRI).



\appendix

\section{Current-density profile in the mixed state}\label{app:current_density}

We derive the current-density profile in the mixed state and obtain the
relevant expressions for the vortex-box geometries that are relevant to Sec.\
\ref{sec:with-transport}. We consider positive magnetic fields $H$, such that
vortices accumulate towards the right of the superconductor when $I>0$. We
first consider the situation where the superconductor is prepared at vanishing
bias current $I=0$, as discussed in Sec.\ \ref{ref:prepared_at_0} and
subsequently analyse the scenario where the bias current $I$ is decreased from
an $I_c(H)$, see Sec.\ \ref{ref:prepared_at_I_c}.

\subsection{Increasing current from $I=0$}\label{app:from_0}

\subsubsection{Fields $H\leq H_s$}
The current-density profiles for states prepared from an initial configuration
with $I=0$ are shown in Fig.\ \ref{fig:current_density}.  For small fields $H
\leq H^\ast$, the superconductor remains in the Meissner state for all values
of the current $I$ and no hysteretic effect is present.  For larger fields
$H^\ast \leq H \leq H_s$, the superconductor is in the Meissner state when $I
\leq I_s(H)$, see Eq.\ \eqref{eq:I_s_vs_H}, and enters the mixed state at
larger currents when $i(-W/2) = i_s$ and vortices penetrate from the left edge
of the film.  For $I > I_s(H)$, the current-density profile at the left of the
film drops linearly from $i_s$,
\begin{equation}\label{eq:mixed_state_right}
   i(x) = i_s - \frac{cd\,H}{4\pi\lambda^2}\left(x + \frac{W}{2}\right),
   \quad -W/2 \leq x \leq x_0,
\end{equation}
and pushes the vortices to the right; these accumulate in a box
$x\in\left[x_0,x_1\right]$ with constant current- and vortex densities $i(x) =
i_p$ and $n(x) =H/\Phi_0$. The expression for the left boundary $x_0$ of the
vortex box is found by equating $i(x)$ in \eqref{eq:mixed_state_right} to
$i_p$ and we find
\begin{equation}\label{eq:x_right}
   x_0(H) = -\frac{W}{2}\left(1 - \frac{2H^\ast}{H}\right),
\end{equation}
independent of $I$ and identical to $-x_o(H)$ in Eq.\ \eqref{eq:x_m_x_M} with
$H_s-H_p$ rewritten as $2\Hast$, see Eq.\ \eqref{eq:H_ast}; the result is
applicable in the wide field interval $|H|\in\left[H^\ast,H_{c2}\right]$ where
it takes values between $W/2$ and $-W/2 + \xi$.

The vortex box extends between $x_0$ and $x_1$; to the right of the box, $x >
x_1$, the current density profile again drops linearly from $i_p$,
\begin{equation}\label{eq:mixed_state_left}
   i(x) = i_p - \frac{cd\,H}{4\pi\lambda^2}\left(x - x_1\right),
   \quad x_1 \leq x \leq W/2,
\end{equation}
and the constraint of fixed bias current $I\! =\!\int\! i(x)\mathrm{d}x$ fixes
$x_1$, that reads
\begin{equation}\label{eq:x_left_A}
   x_1(H,I > I_s) \! = \! W \! \left[\frac{1}{2} \! - \! \sqrt{\frac{H_p}{H} 
   \left(\! 1 \! + \! \frac{(H^\ast)^2}{H H_p}\right)
   \! - \! \frac{I}{I_p}} \right],
\end{equation}
or, after using equation \eqref{eq:I_c_mixed_state} for $I_c(H)$,
\begin{equation}\label{eq:x_left_Abis}
   x_1(H,I > I_s) \!=\! W\left[\frac{1}{2} \!-\! \sqrt{
   \frac{H_p}{H}\frac{I_c(H) - I}{I_p}} \,\right],
\end{equation}
which touches $W/2$ when $I = I_c(H)$.  The left and right boundaries
\eqref{eq:x_right} and \eqref{eq:x_left_A} are shown as black dots in Fig.\
\ref{fig:current_density}(d).

\subsubsection{Fields $H > H_s$}
For larger fields $|H| > H_s$ and vanishing bias current $I = 0$, two
symmetric vortex boxes (where $i = \pm i_p$) are present inside the
superconductor, see Fig.\ \ref{fig:crit_state}(c), the left
(right) one extending from $x_0 = - x_o$ to $x_1 = - x_i$ ($x_2 = x_i $ to
$x_3 = x_o$), see Eq.\ \eqref{eq:x_m_x_M}.  Upon increasing the bias current
$I$, the left box expands to the right as new vortices enter from $-W/2$,
while the right one shifts towards $W/2$ with conserved vortex number as the
vortices therein are trapped.

To find the positions of the four edges of the vortex boxes, we impose that
the current-density profile $i(x)$ (involving linear and constant portions as
follows from Eq.\ \eqref{eq:maxwell_london}) is continuous and the total
transport current is fixed to $I = \int\! \mathrm{d}x \, i(x)$. 
The condition $i(-W/2)=i_s$ fixes
\begin{equation}\label{eq:x_right_bis}
   x_0(H) = - \frac{W}{2}\left(1 - \frac{2H^\ast}{H}\right)
\end{equation}
and the conserved vortex number in the right box, $x_3(I,H) -
x_2(I,H) = x_o(H)-x_i(H)$ together with Eq.\
\eqref{eq:x_m_x_M} requires that
\begin{equation}\label{eq:x_tildes}
	x_3(I,H) = x_2(I,H) + \frac{W}{2}\left(1 - \frac{H_s}{H}\right).
\end{equation}
Finally, the linear portion of $i(x)$ between $x_1$ and $x_2$ connects
the critical current densities $i_p$ and $-i_p$, hence
\begin{align}\label{eq:x_<_app}
   x_2(I,H) &= x_1(I,H) + \frac{W}{2} \frac{2H_p}{H}
\end{align}
which, when replaced to Eq.\ \eqref{eq:x_tildes}, yields
\begin{equation}
   x_3(I,H) = x_1(I,H) + \frac{W}{2} \left(1 - \frac{2H^\ast}{H}
		 +\frac{H_p}{H}\right).
\end{equation}
Integrating the profile $i(x)$, we find the total current
\begin{eqnarray}\nonumber
   I &= \displaystyle{i_p\left(x_1 + x_2\right) 
   + \frac{1}{2}\frac{cdH}{4\pi\lambda^2}
   \left(x_0 + \frac{W}{2}\right)^2}
    \\
   \label{eq:total_current_two_boxes}
   &\qquad\quad \displaystyle{
   - \frac{1}{2}\frac{cdH}{4\pi\lambda^2}
   \left(\frac{W}{2} - x_3\right)^2}.
\end{eqnarray}

The above equations \eqref{eq:x_right_bis}--\eqref{eq:total_current_two_boxes}
determine the four box boundaries $x_0,x_1,x_2,x_3$ as functions of given
field $H$ and current $I$; they apply to the regime $I < I_t(H)$, see Eq.\
\eqref{eq:I_t_vs_H}, where the right box is separated from the right boundary,
$x_3 < W/2$.  Inserting $x_0$ from \eqref{eq:x_right_bis} and $x_3 = W/2$ into
\eqref{eq:total_current_two_boxes}, we find the current $I_t(H)$ as given in
\eqref{eq:I_t_vs_H}. This result for $I_t(H)$ then allows us to write the
results for the box boundaries in a compact form,
\begin{equation}\label{eq:x_LL}
   x_3= \frac{W}{2}\left(1+ \frac{2H_p}{H} [1-S(H,I)]\right),
\end{equation}
\begin{equation}\label{eq:x_LR}
   x_2 =  \frac{W}{2}\left(\frac{H_s}{H} 
   +\frac{2H_p}{H} [1-S(H,I)]\right),
\end{equation}
and
\begin{equation}\label{eq:x_RL}
   x_1 =  \frac{W}{2}\left(\frac{H_s}{H}
   - \frac{2H_p}{H} S(H,I)\right),
\end{equation}
with
\begin{equation}\label{eq:S_app}
   S(H,I) =  \sqrt{1 + \frac{H}{H_p} 
   \frac{\left[I_t(H) - I\right]}{I_p}}.
\end{equation}

The corresponding current-density profile is shown in blue in Fig.\
\ref{fig:current_density}(e) with the edges of the left and right vortex boxes
marked by black dots.  For $I = I_t(H)$, we have $S=1$ and $x_3 =W/2$.

For larger currents, vortices start exiting the right box which gradually shrinks. 
In this situation, the expression for the total current $I$ simplifies to 
\begin{equation}\label{eq:total_current_one_box}
   I = i_p\left(x_1 + x_2\right) 
   + \frac{1}{2}\frac{cdH}{4\pi\lambda^2}
   \left(x_0 + \frac{W}{2}\right)^2
\end{equation}
and using the relations \eqref{eq:x_right_bis} and \eqref{eq:x_<_app}, the new
edges of the vortex boxes are found to be
\begin{equation}\label{eq:x_LR_bis}
   x_2 = \frac{W}{2}\left(1 - \frac{I_d(H) - I}{I_p}\right)
\end{equation}
and
\begin{equation}\label{eq:x_RL_bis}
   x_1 = \frac{W}{2}\left(1 - \frac{2H_p}{H} - \frac{I_d(H) - I}{I_p}\right),
\end{equation}
Similar as above, we have made use of the expression \eqref{eq:I_d_vs_H} for
$I_d(H)$, which is to be obtained by the condition $x_2 = W/2$ for the
complete disappearance of the right vortex box and inserting the expressions
for $x_0$ and $x_1$ into \eqref{eq:total_current_one_box}.  The
current-density profile for $I_t(H)<I\leq I_d(H)$ is shown in magenta in Fig.\
\ref{fig:current_density}(e).  Finally, when $I = I_d(H)$, the right vortex
box disappears as $x_2 = W/2$, and the mixed state displays a single vortex
box described by Eqs.\ \eqref{eq:x_right} and \eqref{eq:x_left_Abis} above.

\subsection{Decreasing current from $I_c(H)$}\label{app:from_I_c}

The presence of vortices leads to hysteretic effects, changing the film's
behavior upon decreasing the bias current $I$ from $I_c(H)$ as compared to
increasing $I$ from zero.

After the state preparation at $I$ slightly below $I_c(H)$ for fields
$H>\Hast$, vortices remain trapped in their position as long as the absolute
value of the local current density does not exceed $i_p$.  Upon decreasing the
bias current $I$, the current distribution is therefore shifted rigidly
towards smaller values, see the upper black and red curves in Fig.\
\ref{fig:current_density_rev}(b), until the current density in the vortex box
reaches $-i_p$ --- this happens when $I=I_c(H) - 2I_p \equiv I_r(H)$, see the
black dashed line in (a).  Beyond this point, vortices start rearranging and
the vortex box shifts to the left, while maintaining its width and total
vortex number.  Correspondingly, a new vortex-free region with linear
current-density profile  is formed close to the right edge (lower red curve in
(b)).  In this regime, the edges of the vortex box are found by imposing that
the current-density profile is continuous and carries the total current
\begin{eqnarray}\label{eq:I_01}
   I =& -i_pW + \displaystyle{\frac{1}{2}\frac{c\, dH}{4\pi\lambda^2}
   \left(x_0 + \frac{W}{2}\right)^2} \\
   & \qquad \qquad \displaystyle{- \frac{1}{2}\frac{c\, dH}{4\pi\lambda^2}
   \left(\frac{W}{2} - x_1\right)^2.}\nonumber
\end{eqnarray}
The distance $x_1 - x_0$ does not depend on $I$ and its evaluation at
$I_c(H)$ (where $x_0$ is given by \eqref{eq:x_right_bis} and $x_1 = W/2$)
produces the condition $x_1 - x_0 = W(1 - H^\ast/H)$. Combining this result with
Eq.\ \eqref{eq:I_01}, we find that
\begin{equation}\label{eq:x_>_app_I_c}
   x_0 = -\frac{W}{2}\left(1-\frac{2H^\ast}{H}+\frac{H_p}{H^\ast}
   \frac{\left[I_r - I\right]}{I_p} \right)
\end{equation}
and
\begin{equation}\label{eq:x_<_app_I_c}
   x_1 = \frac{W}{2}\left(1-\frac{H_p}{H^\ast}
   \frac{\left[I_r-I\right]}{I_p} \right).
\end{equation}
The edges of the vortex box in the regime $I<I_c(H) - 2I_p = I_r$ are marked with
black dots in Fig.\ \ref{fig:current_density_rev}(b).  Finally, when $I
=-I_c(H)$, the vortex box touches the left edge of the sample, while $i(W/2)$
reaches $-i_s$, and the system has lost any memory of the preparation at
$I_c(H)$.

\section{Magnetic moment for a 2D thin film superconductor}\label{app:m_moments}

Sweeping the magnetic field down from a reversal value $H_0 >H_s$, the current
distribution across the thin film depends on the vortex density via the
Maxwell-London equation \eqref{eq:maxwell_london}, with the latter to be
determined by the interplay of surface and bulk pinning.  Along the magnetic
loop, both the current- and vortex profiles change shape as shown in Fig.\
\ref{fig:m_current_density} with separate colors. Inserting the
current-density profiles $i(x)$ into the definition
\eqref{eq:magnetization_def_film} for the magnetic moment, we obtain $m(H)$
for the various regimes, as plotted in Fig.\ \ref{fig:m_moment} using the same
color-coding.  In this appendix, we present the analytical expressions for the
current- and vortex densities and for the magnetic moments as a function of
the magnetic field, thus complementing the discussion in Sec.\
\ref{sec:m_decreasing_fields} with quantitative results.

\subsubsection{Fields $H_1 \leq H\leq H_0$}

As the magnetic field is reversed at $H_0$, vortices stay trapped in their
original positions $|x|\in\left[x_i(H_0), x_o(H_0)\right]$, see Eqs.\
\eqref{eq:x_m_x_M}, and the vortex distribution is given by
\begin{equation}\label{eq:ny_H1}
   n(x>0) = \begin{dcases} 
		0, \qquad\qquad x\leq x_i(H_0),\\
                ~~H_0/\Phi_0, \quad x \in \left[x_i(H_0), x_o(H_0)\right],\\
		0, \qquad\qquad x > x_o(H_0).
         \end{dcases}
\end{equation}
\vspace{1pt}
and symmetrically for $x < 0$. This produces the (antisymmetric in $x$)
current distribution
\onecolumngrid
\begin{equation}\label{eq:iy_H1}
   i(x>0) =  -\frac{cd}{4\pi\lambda^2} \begin{dcases} H x,
                            \qquad\qquad\qquad\qquad \qquad\quad ~x\leq x_i(H_0),\\
                            H x - H_0\, [x - x_i(H_0)],
                            \qquad\quad ~x_i(H_0) \leq x \leq x_o(H_0),\\
                            Hx - H_0\, [x_o(H_0) - x_i(H_0)],
                            \quad x > x_o(H_0),
         \end{dcases}
\end{equation}
that involves three linear in $x$ segments as shown with the blue line in Fig.\
\ref{fig:m_current_density}.  Integrating the current-density profile as
prescribed in Eq.\ \eqref{eq:magnetization_def_film} (see also Eq.\
\eqref{eq:xi_int}), we obtain the magnetic moment (shown in blue in the
magnetic loop in Fig.\ \ref{fig:m_moment})
\begin{align}\label{eq:m(H)_abv_H1}
   m(H) &=-\frac{dWL}{48\pi}\frac{W^2}{\lambda^2} 
       \left[H  
                 + \frac{H_0}{2} \left(1 - \frac{2H^\ast}{H_0}\right)^3 
                 - \frac{H_0}{2} \left(\frac{H_p}{H_0}\right)^3
                            -\frac{3}{2}\left(H_0 - H_s\right) \right].
\end{align}

\twocolumngrid

With reference to the piecewise current-density integral, cf.\
\eqref{eq:xi_int}, the first and last terms in \eqref{eq:m(H)_abv_H1}
originate from the film edge at $W/2$, while the second and third term derive
from the vortex box boundaries at $x_o(H_0)$ and $x_i(H_0)$, respectively, see
Eq.\ \eqref{eq:x_m_x_M} and use $H_s-H_p = 2 H^\ast$.  Making use of the
result \eqref{eq:m_mom_mixed} for the mixed-phase moment $m(H>H_s)$ in the
main text, we can reexpress the above result in the form
\begin{align}\label{eq:m(H)_abv_H1_1}
   m(H) &= m(H_0) - \frac{dWL}{48\pi}\frac{W^2}{\lambda^2}(H-H_0)
\end{align}
to find that the Meissner slope is retraced downwards from the point
$(H_0,~m(H_0))$ in the $m$--$H$ diagram. This result can be easily understood
by inspection of the current-density profile \eqref{eq:iy_H1} (see also Fig.\
\ref{fig:m_current_density}(a)): although vortices are present, they are
pinned with a fixed density $n_0 = H_0/\Phi_0$ and box boundaries $x_i(H_0)$
and $x_o(H_0)$ that do not depend on $H$.  As a consequence, the derivative
$\partial_H i = -(cd/4\pi\lambda^2) \, x$ extends uniformly over the entire
width of the film, see Eqs.\ \eqref{eq:maxwell_london} and \eqref{eq:london},
that results in a Meissner response.

The vortex distribution changes when the outermost vortex can move again, i.e., for
\begin{equation}
   i(\pm x_o(H_0), H_1)=\pm i_p 
\end{equation}
from which we find the field boundary $H_1$,
\begin{equation}
   H_1 = H_0 \frac{H_0-H_s - H_p}{H_0 - H_s + H_p}.
\end{equation}

\subsubsection{Fields $H_2 \leq H\leq H_1$}

Lowering the field below $H_1$, the outermost vortices rearrange themselves to define
a new density profile
\begin{equation}\label{eq:ny_H2}
   n(x>0) = \begin{dcases} 
                0, \qquad\qquad x\leq x_i(H_0),\\
		H_0/\Phi_0, \quad ~~x\in\left[x_i(H_0), x_s(H)\right],\\
                H/\Phi_0, \qquad x\in\left[x_s(H), x_o(H)\right],\\
                0, \qquad\qquad x > x_o(H),
         \end{dcases}
\end{equation}
and the current distribution reads (we express the bulk critical current $i_p$
by $H_p$, $i_p = (cd/4\pi\lambda^2)\, H_p\, W/2$)

\onecolumngrid

\begin{equation}\label{eq:iy_H2}
   i(x>0) = -\frac{cd}{4\pi\lambda^2}
         \begin{dcases} 
                   H x, \qquad\qquad\qquad\qquad\qquad\quad x\leq x_i(H_0),\\
                   H x - H_0 \, [x - x_i(H_0)], \qquad\quad
                   x_i(H_0)\leq x\leq x_s(H) ,\\
                   -H_p\, W/2, ~\qquad\qquad\qquad\qquad
                   x_s(H)\leq x\leq x_o(H),\\
                   -H_p \, W/2 + H \,[x - x_o(H)],
                   \quad x > x_o(H).
         \end{dcases}
\end{equation}
The edges $x_s(H)$ and $x_o(H)$ of the region where vortices redistribute are
found by imposing the condition $i(x_s(H)) =  i_p$ together with the
conservation of vortex number \eqref{eq:n_vortices_cons}, see Eqs.\
\eqref{eq:y_<} and \eqref{eq:y_>}.  Together, these conditions provide the
results \eqref{eq:y_<} and \eqref{eq:y_>} in the main text.  At the field $H_2
= H_0 - H_s - H_p$, cf. Eq.\ \eqref{eq:H_2}, the outer edge reaches the film
surface, $x_o(H_2) = W/2$, and vortices start exiting the sample.  The
current- and vortex distributions for fields $H_2 \leq H \leq H_1$ are shown
in orange in Fig.\ \ref{fig:m_current_density} and give rise a nonlinear
dependence of the magnetic moment on $H$ (shown in orange in Fig.\
\ref{fig:m_moment}),
\begin{align}\label{eq:m(H)_abv_H2}
   m(H)&= - \frac{dWL}{48\pi}\frac{W^2}{\lambda^2} 
   \left[ H + \frac{H}{2}\left(\frac{H_0-H_s-H_p}{H}\right)^3 
            + H_p \left(\frac{2H_p}{H_0-H}\right)^2
            -\frac{H_0}{2}\left(\frac{H_p}{H_0}\right)^3
            - \frac{3}{2}\left(H_0-H_s\right)\right].
\end{align}
Again, the first and last terms originate from the film edge $W/2$, the second
and third terms derive from $x_o(H)$ and $x_s(H)$ (note that $x_o(H_0)$ as
given by Eq.\ \eqref{eq:x_m_x_M} changes to $x_o(H)$ given by Eq.\
\eqref{eq:y_>} when the box splits), respectively, and the fourth term comes
from the inner boundary $x_i(H_0)$. This association of terms with boundaries
will repeat itself also in the expressions below. Furthermore, we note that
$\partial_H m$ is continuous at $H_1$: indeed, the derivatives $\partial_H i$
of Eqs.\ \eqref{eq:iy_H1} and \eqref{eq:iy_H2} match up, as the interval
$[x_s(H),x_o(H)]$ starts appearing only when $H$ drops below $H_1$ and $H
x_o(H) = (H_0-H_s-H_p)\, W/2$ is independent of $H$ according to Eq.\
\eqref{eq:y_>}.

\subsubsection{Fields $0 \leq H\leq H_2$}

Decreasing the magnetic field below $H_2$, the total vortex number is not
conserved as the outermost vortices gradually leave the sample,
\begin{equation}\label{eq:ny_0}
   n(x >0) = \begin{dcases} 
                        0, \qquad\quad ~x\leq x_i(H_0),\\
                        H_0/\Phi_0, \quad
                        x\in\left[x_i(H_0), x_s(H)\right],\\
                        H/\Phi_0, \quad ~x\in\left[x_s(H), W/2\right],
         \end{dcases}
\end{equation}
with the corresponding critical current densities $i(x) = \pm i_p$ extending to
the sample edges. The full $i(x)$-profile reads
\begin{equation}\label{eq:iy_0}
   i(x>0) =  -\frac{cd}{4\pi\lambda^2} \begin{dcases} 
		H x, \qquad\qquad\qquad\qquad\quad
                x \leq x_i(H_0),\\
                H x - H_0 \,[x-x_o(H_0)],\quad
                x_i(H_0)\leq x\leq x_s(H),\\
                 -H_p\, W/2, \qquad\qquad\qquad ~x > x_s(H),\\
         \end{dcases}
\end{equation}
with $x_s(H)$ still given by Eq.\ \eqref{eq:y_<}. Again, the derivatives
$\partial_H i$ of Eqs.\ \eqref{eq:iy_H2} and \eqref{eq:iy_0} match up at
$H_2$: since $x_o(H) = W/2$ at $H_2$, the vortex free region between $x_o$ and
$W/2$ disappears and the last region $x > x_o(H)$ of Eq.\ 
\eqref{eq:iy_H2} is absent. As a result, the derivative $\partial_H m$ is
continuous at $H_2$.  This profile remains valid until all vortices in
the outermost region $|x|\geq x_s(H)$ have left the film, i.e., until $H = 0$.
The vortex- and current density profiles in the field interval $0\leq H \leq
H_2$ are shown in green in Fig.\ \ref{fig:m_current_density}. As the current
density is constant and equal to $\pm i_p$ in a large part of the sample, the
magnetic moment depends only weakly on the field in this regime,
\begin{align}\label{eq:m(H)_abv_0}
   m(H) = -\frac{dWL}{48\pi}\frac{W^2}{\lambda^2}
   \left[H_p\left(\frac{2H_p}{H_0 - H}\right)^2 
   -\frac{H_0}{2}\left(\frac{H_p}{H_0}\right)^3 - \frac{3}{2} H_p\right],
\end{align}
as can be seen by the green curve in Fig.\ \ref{fig:m_moment}. Since the slope
in $i(x)$ at $W/2$ is absent, see the green line in Fig.\
\ref{fig:m_current_density} left, there is no linear in $H$ term in Eq.\
\eqref{eq:m(H)_abv_0}; furthermore, the boundary $x_o(H)$ is absent in this
field interval.  As discussed in the main text, the above discussion for
fields in the interval $0 \leq H \leq H_1$ is valid as long as $H_1 \geq 0$,
i.e., when $H_0\geq H_s + H_p$.  Else, the field $H_1 < 0$ turns negative and
the outermost vortices are expelled from the film as soon as they can move
away from their pins.

\subsubsection{Fields $H_3 \leq H\leq 0$}

Decreasing the field further below $\min(0, H_1)$, the vortex region shrinks
together with $x_s(H)$ and the outermost region turns vortex free,
\begin{equation}\label{eq:ny_H3}
   n(x > 0) = \begin{dcases} 
                     0, \qquad\quad ~x\leq x_i(H_0),\\
                     H_0/\Phi_0, \quad x\in\left[x_i(H_0), x_s(H)\right],\\
                     0, \qquad\quad ~x > x_s(H).
         \end{dcases}
\end{equation}
Correspondingly, the current density is once again linear close to the edges
of the sample, with the overall profile taking the form
\begin{equation}\label{eq:iy_H3}
   i(x>0) = -\frac{cd}{4\pi\lambda^2}
                 \begin{dcases} 
		H x, \qquad\qquad\qquad\qquad\qquad\quad ~x\leq x_i(H_0),\\
                H x - H_0\,[x - x_i(H_0)], \qquad\quad
                ~x_i(H_0) \leq x\leq x_s(H),\\
                -H_p\, W/2 + H\,[x - x_s(H)],
                \quad ~x > x_s(H).
         \end{dcases}
\end{equation}
These results are valid as long as $i(W/2)\leq i_s$, prohibiting anti-vortices
to enter the film, that defines the field $H_3$ as given by Eq.\
\eqref{eq:H_3}.  The corresponding vortex- and current densities are shown in
red in Fig.\ \ref{fig:m_current_density} and give rise to an
approximately linear dependence of the magnetic moment on the field $H$
(see the red segments in Fig.\ \ref{fig:m_moment}),
\begin{align}\label{eq:m(H)_abv_H3}
   m(H)= - \frac{dWL}{48\pi}\frac{W^2}{\lambda^2}
   \left[H + \frac{H_0}{2}\left(\frac{2H_p}{H_0 - H}\right)^3
            - \frac{3H_0}{2} \frac{2H_p}{H_0-H}
            -\frac{H_0}{2}\left(\frac{H_p}{H_0}\right)^3
            + \frac{3H_p}{2}\right].
\end{align}
Note, however, that the Meissner slope is spoiled by the field dependence of
$x_s(H)$ in \eqref{eq:iy_H3}, cf.\ Eq. \eqref{eq:iy_H1} where no such
dependence of the vortex density $n(x)$ on $H$ shows up; furthermore the
term $H x_s(H)$ is responsible for the deviation from the linear dependence of
$m(H)$.  The association of terms and boundaries in \eqref{eq:m(H)_abv_H3}
now involves three terms from $x_s(H)$ and the $x_o(H)$-term is absent.

\subsubsection{Fields $-H_0 \leq H\leq H_3$}

When anti-vortices start penetrating into the thin film for $H\leq H_3$,
neighboring vortex- and anti-vortex domains form (if pinning is sufficiently
strong), and the vortex distribution reads
\begin{equation}\label{eq:ny_H4}
   n(x>0) = \begin{dcases} 
                 0, \qquad\quad ~x \leq x_i(H_0),\\
                 H_0/\Phi_0, \quad x\in\left[x_i(H_0), x_f(H)\right],\\
                 H/\Phi_0, \quad ~x\in\left[x_f(H), x_o(H)\right],\\
                 0, \qquad\quad ~x > x_o(H),
         \end{dcases}
\end{equation}
where the density $n(x) = H/\Phi_0$ is negative in the region $x_f(H)\leq
|x|\leq x_o(H)$, $x_o(H) = (W/2)[1+2\Hast/H)$ follows from \eqref{eq:x_m_x_M} with $H <0$,
and the position of the vortex--anti-vortex front $x_f(H)$ is found the same
way as the step position $x_s(H)$ before, i.e., as the point where $i(x)$
crosses $i_p$, hence $x_f(H) = (W/2)[2H_p/(H_0-H)]$, see Eq.\ \eqref{eq:y_<}.
The current distribution reads
\begin{equation}\label{eq:iy_H4}
   i(x>0) = -\frac{cd}{4\pi\lambda^2} \begin{dcases} 
       H x, \qquad\qquad\qquad\qquad\qquad\quad x\leq x_i(H_0),\\
       H x - H_0\,[x - x_i(H_0)],
       \qquad\quad x_i(H_0)\leq x \leq x_f(H),\\
       -H_p\, W/2, \qquad\qquad\qquad\qquad ~x_f(H)\leq x \leq x_o(H),\\
       -H_p\, W/2 + H\, [x - x_o(H)], 
       \quad  x  > x_o(H),\\
         \end{dcases}
\end{equation}
and is analogous to Eq.\ \eqref{eq:iy_H2} with the replacement of $x_s(H)\to
x_f(H)$.  As the field is reduced, vortices at $x_f(H)$ annihilate with
anti-vortices until $x_f(H) = x_i(H_0)$ and only anti-vortices are left. As
discussed in the main text, this happens for $H = -H_0$.
The vortex- and current densities for fields $-H_0 \leq H \leq H_3$ are shown
in purple in Fig.\ \ref{fig:m_current_density} and give rise to the magnetic
moment (see purple segments in Fig.\ \ref{fig:m_moment})
\begin{align}\label{eq:m(H)_abv_H4}
   m(H)&=- \frac{dWL}{48\pi}\frac{W^2}{\lambda^2}
   \left[-\frac{H}{2} + \frac{H}{2}\left(1+\frac{2\Hast}{H}\right)^3
            + H_p \left(\frac{2H_p}{H_0-H}\right)^2
            -\frac{H_0}{2}\left(\frac{H_p}{H_0}\right)^3
            - \frac{3H_s}{2}\right].
\end{align}

\twocolumngrid

\bibliography{refs_film}
\end{document}